\documentclass[aps,prb,twocolumn,amsmath,amssymb,unsortedaddress]{revtex4-1}

\usepackage{graphicx}
\usepackage{dcolumn}
\usepackage{bm}
\usepackage{url}
\usepackage[colorlinks=true,linkcolor=blue,citecolor=blue,urlcolor=blue]{hyperref}

\usepackage[normalem]{ulem}

\usepackage{color}

\begin{document}

\title{Phase Stability and Properties of Manganese Oxide Polymorphs: \\
Assessment and Insights from Diffusion Monte Carlo}

\author{Joshua A. Schiller$^1$}
\author{Lucas K. Wagner$^2$}
\author{Elif Ertekin$^{1,3}$}
\affiliation{$^1$Department of Mechanical Science \& Engineering, 1206 W Green Street, University of Illinois at Urbana-Champaign, Urbana IL 61801}
\affiliation{$^2$Department of Physics, University of Illinois at Urbana-Champaign, Urbana IL 61801}
\affiliation{$^3$International Institute for Carbon Neutral Energy Research (WPI-I$^2$CNER), Kyushu University, 744 Moto-oka, Nishi-ku, Fukuoka 819-0395, Japan}

\email[e-mail: ]{ertekin@illinois.edu}

\date{\today}

\begin{abstract} 
\noindent We present an analysis of the polymorphic energy ordering and properties of the rock salt and zincblende structures of manganese oxide using fixed node diffusion Monte Carlo (DMC). Manganese oxide is a correlated, antiferromagnetic material that has proven to be challenging to model from first principles across a variety of approaches. Unlike conventional density functional theory and some hybrid functionals, fixed node diffusion Monte Carlo finds the rock salt structure to be more stable than the zincblende structure, and thus recovers the correct energy ordering.  Analysis of the site-resolved charge fluctuations of the wave functions according to DMC and other electronic structure descriptions give insights into elements that are missing in other theories. While the calculated band gaps within DMC are in agreement with predictions that the zincblende polymorph has a lower band gap, the gaps themselves overestimate reported experimental values. 
\end{abstract}


\maketitle

\section{Introduction} 

Transition-metal oxides exhibit a rich variety of intriguing phenomena, including metal-to-insulator transitions\cite{Mott1990,Morin1959}, high-temperature superconductivity\cite{Bednorz1986,Kamihara2008}, colossal magnetoresistance\cite{Urushibara1995}, and colossal dielectric constants\cite{Lunkenheimer2010}.  These properties are closely related to the strongly correlated nature of the localized $d$-orbital electrons.  The presence of electron correlations also render transition metal oxides very challenging to model from first principles. One classic example of a correlated metal oxide is manganese oxide (MnO), which is of interest for several potential applications including solar energy conversion \cite{Peng2012}, photoelectrochemical water splitting \cite{Kanan2012,Toroker2013}, and as a magneto-piezoelectric semiconductor \cite{Gopal2004}. While the ground state of MnO is rock salt (RS), a few years ago metastable wurtzite (WZ) was grown by thermal decomposition on a carbon template \cite{MinNam}. More recently, predictions of a reduced band gap and favorable hole transport properties in the wurtzite phase \cite{Peng2012,Peng2013} have been verified in experiment \cite{PhysRevX.5.021016}. 

Manganese oxide possesses a $d^5$ electronic structure, and exhibits antiferromagnetic ordering of the Mn atoms. Several recent first-principles studies have explored in detail the properties of two polymorphs of MnO: rock salt (RS) and zincblende (ZB)\cite{Peng2013,Schron2010,Franchini2005,Towler1994}. Within the RS and ZB polymorphs, there is antiferromagnetic ordering along the [111] (AF2) and [001] (AF1) directions, respectively\cite{Schron2010}. Although rock salt is the ground state structure, the generalized gradient approximation (GGA) and some hybrid functionals (HSE06) erroneously predict that the zincblende structure is lower in energy\cite{Peng2013,Schron2010}. 
The failure of conventional DFT and even some hybrids to obtain the correct energetic ordering shows the importance of correlation in the phase stability of these materials. The challenges in describing correlated materials within DFT arise from its approximate treatment of electron correlation and exchange. This is true whether one uses the local density approximation\cite{MartinText}, the generalized gradient approximation\cite{MartinText}, or a hybrid functional\cite{Adamo1999,becke1993new,heyd2003hybrid}. 
The accuracy and transferability of a given approximation across a spectrum of materials, or even for different polymorphs of the same material, must ultimately be justified {\it a posteriori}, by comparison to experiment.  

By contrast, in this work we use fixed-node diffusion Monte Carlo (FN-DMC), a type of quantum Monte Carlo method, to assess the properties of the zincblende and the rock salt polymorphs of MnO. The reasons are twofold.  First, in quantum Monte Carlo, statistical sampling is used to approximate the many-body wave function and evaluate total energies directly from the first-principles many-body Schr\"{o}dinger equation, greatly reducing the extent of approximation necessary.  Quantum Monte Carlo methods therefore offer a parameter-free, systematically improvable approach. Because of their direct treatment of electron correlation, they are amongst the most accurate electronic structure approaches available today~\cite{Petruzielo:2012bl,Foulkes:2001td,Grossman01}.  
Second, although the FN-DMC method is in principle exact when the nodal structure is exactly known, there remain outstanding questions as to the practical accuracy of the technique.  It is important to test how well simple nodal surface do in practice. This material system offers a test of the capabilities, using ``best practices" for DMC simulation of solids as they are currently understood, to obtain quantitative descriptions of challenging correlated oxide materials. 

For MnO, our results show that the DMC method obtains accurate descriptions of the ground state of both the RS and ZB phase, including their relative energies and lattice constants.  We find that the magnitude of the fixed node error for the ground state is small in comparison to the substantial improvement that comes from adopting an explicitly correlated approach.  Because DMC samples the true many-body wave function, we analyze the properties of this highly accurate model to assess the physical reason for the failure of DFT methods. Thus, the use of FN-DMC helps to reveal aspects, such as charge fluctuation and localization, that 
may not be accurately captured by other methods.  In agreement with other theoretical methods, we also find that the band gap of the ZB phase is substantially lower than that of RS phase according to FN-DMC.  However, FN-DMC overestimates the band gaps of both polymorphs in comparison to experiment.  We discuss possible reasons for the overestimate. 

\begin{figure}[h]
\includegraphics{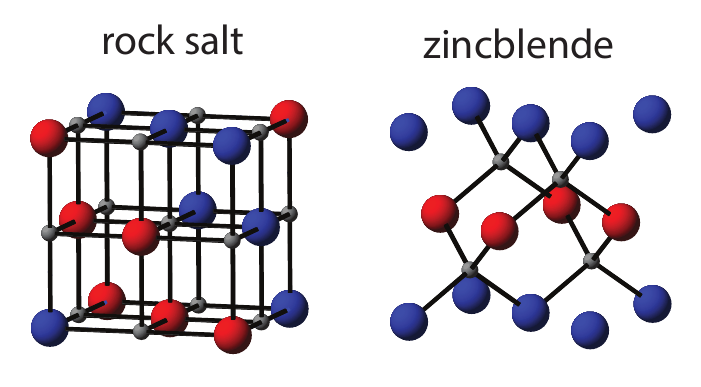}
\caption{\label{atomization} (Color online).  The rock salt (left) and zincblende (right) polymorphs of manganese oxide; both exhibit an antiferromagnetic ordering of Mn atoms. The grey atoms are oxygen and the blue or red are opposing spin manganese atoms. In the rock salt structure each Mn atom possesses a neighboring octahedral field of O atoms; in the zincblende structure the neighboring field of O atoms surrounding each Mn atom is tetrahedral instead. }
\end{figure}

\section{DFT and DMC Methodology}

The QMC calculations reported here were carried out within the FN-DMC framework as implemented in the QWalk code~\cite{Wagner:2009dy}, with single-determinant Slater-Jastrow trial wave functions constructed from DFT Kohn-Sham orbitals, with variance-minimized two-body Jastrow coefficients, and a time step of 0.004 au. We also assessed the sensitivity of the DMC energy to various forms of trial wave functions, such as two and three body Jastrow factors and both energy and variance minimization to optimize the Jastrow parameters. However in all cases we find the DMC energies to be statistically equivalent. This is similar to our observations for DMC simulations of  the wide band gap material zinc oxide, also using small-core BFD pseudopotentials and a similar simulation strategy~\cite{YuWagnerErtekinJCP}.

Ground state energies were determined by twist averaging the DMC energies calculated at real-valued $k$-points, which corresponds to a $2\times2\times2$ grid in each supercell. 
Scalar-relativistic energy-consistent Hartree-Fock pseudopotentials ([Ne] core for Mn) as implemented by Burkatzki, Filippi, and Dolg (BFD) \cite{Burkatzki2007} were used to remove the core electrons. 
These pseudopotentials are designed for use within QMC and there are now several indications in the literature that they are well-suited for DMC simulations of solids\cite{ZhengWagner2015,cerium,Shulenburger2015,KolorencMitasPRL}. The rock salt structure of MnO has previously been studied within DMC\cite{Mitas2010}; to this analysis we now provide a comparison between the ZB and RS polymorphs, physical insights into the electronic structure of the two phases, and statistical analysis of the many body wave functions to reveal the reasons for the failure of conventional and hybrid DFT to obtain the correct energy ordering.

To obtain the trial wave functions for the DMC calculations, we carried out DFT simulations for the RS and ZB phases. For these simulations, we used the CRYSTAL code\cite{dovesi2005crystal} and gaussian-type localized basis sets to expand the Kohn-Sham orbitals. The DFT results presented here implement the ``PBE1$_x$" framework in which the degree of exact exchange mixing $\alpha$ is systematically varied. We do this to study the effect on both the DFT results themselves as well as the the final DMC results arising from different selections of trial wave functions.  

\section{Density Functional Theory Results and Construction of Trial Wave Functions}


\subsection{Effect of Exchange Mixing}

\begin{figure}[h]
\includegraphics[width=8.5cm]{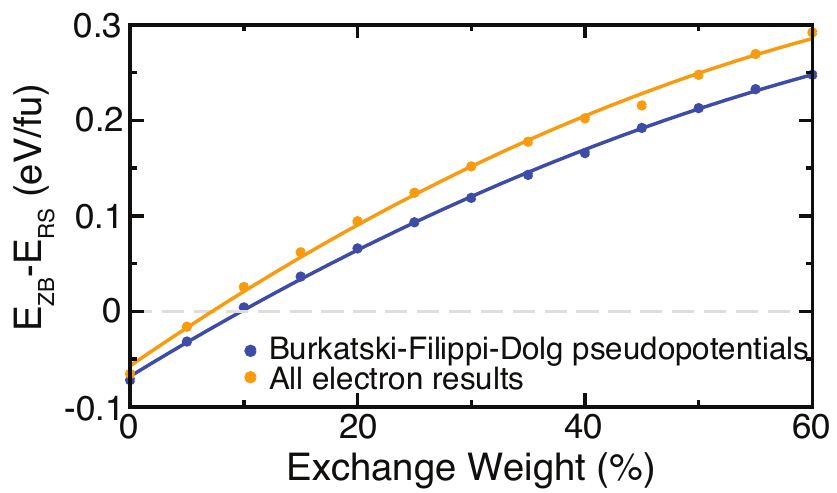} 
\caption{ \label{energyvalpha} (Color online). The energy difference $E_{ZB}-E_{RS}$ according to DFT-PBE1$_x$, obtained from all electron calculations (orange) and with  Burkatski-Filipi-Dolg (Hartree-Fock) pseudopotentials (blue). According to all electron results, for $\alpha = 0$ the ZB phase is lower in energy but as $\alpha$ increases the RS phase becomes stable. The crossing occurs around $\alpha=10$\%. When Burkatski-Filipi-Dolg pseudopotentials are used, the trends are very similar.  }
\end{figure}

\begin{figure*} 
\includegraphics[width=6in]{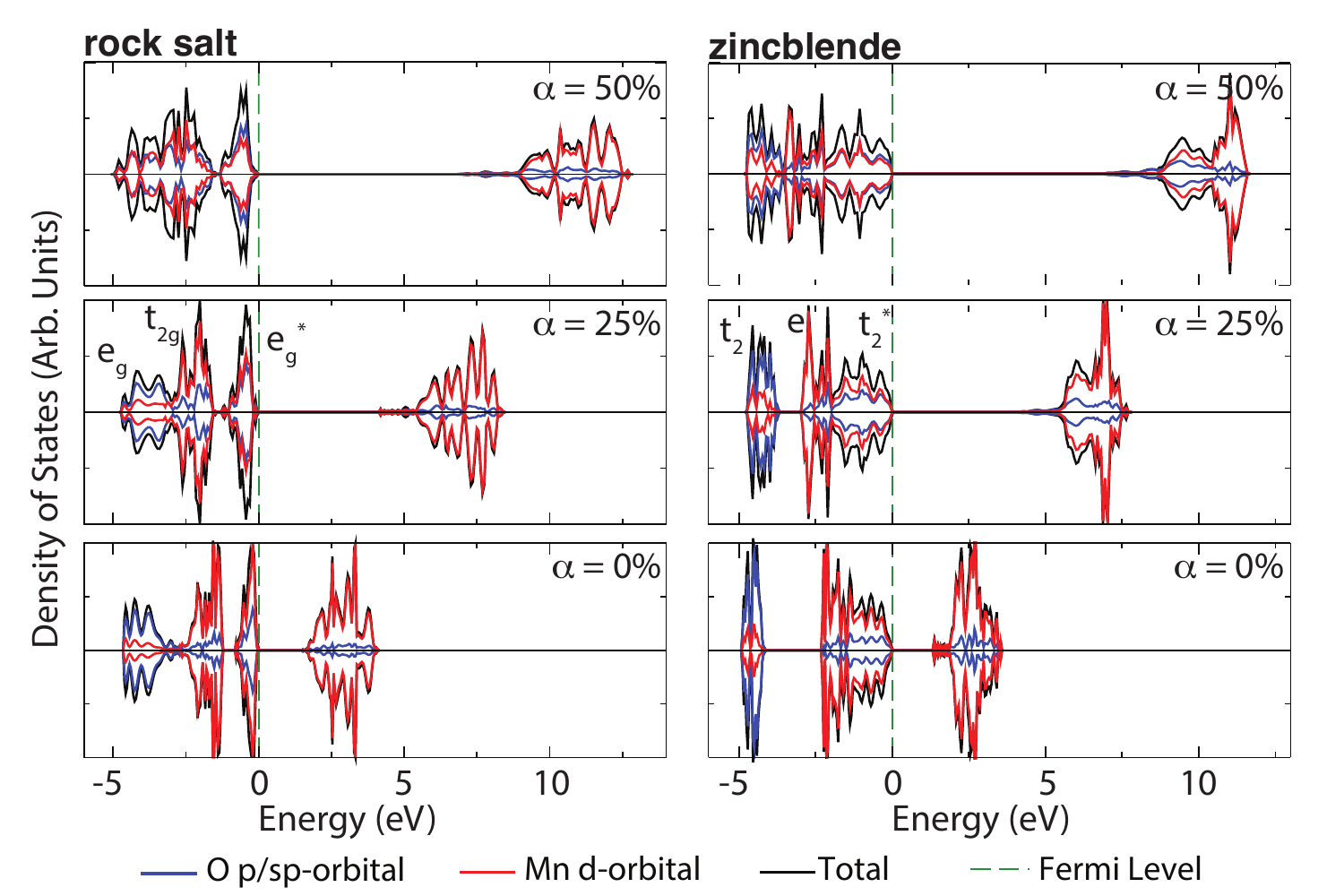}
\caption{\label{pods} (Color online).  The partial density of states for O $2s$,$2p$ and Mn $3d$ orbitals, as obtained within DFT for different degrees of exchange mixing $\alpha$.  As $\alpha$ increases, the band gap becomes larger as expected. Also below the valence band maximum, the $p-d$ hybridization increases as the relative Mn $3d$ orbital energies near the VBM drop.}
\end{figure*}

To begin, we construct a set of trial wave functions for the DMC calculations using the DFT-PBE1$_x$ approach, in which the degree of exchange mixing $\alpha$ is varied between 0 and 60\%.    
In Figure \ref{energyvalpha}, we illustrate the ground state energy difference $(E_{ZB} - E_{RS})$ per formula unit MnO, as a function of the degree of exchange mixing used in the DFT-PBE1$_x$ calculations. 
For comparison, we also show all electron results as well.
It is encouraging that the two sets of results are quite similar, which suggests that the relativistic Hartree-Fock pseudopotentials are not affecting the analysis substantially. 
For these calculations, 4 atom unit cells were used for both RS and ZB in conjunction with an $8\times8\times8$ Monkhorst-Pack sampling of $k$-points in the Brillouin zone. 
The RS lattice constant was set to 4.43 \AA, matching experiment\cite{Roth1958}, while the ZB lattice constant was set to the PBE0  lattice constant of 4.73 {\AA} since the experimental value is not known.  
Positive values of $(E_{ZB} - E_{RS})$ in Fig. \ref{energyvalpha} denote more stable rock salt phase. 

Consistent with previous results\cite{Peng2013,Schron2010}, we find that without exchange mixing ($\alpha = 0$\%, PBE) the ZB phase is more stable (by $\approx$ 70 meV/fu in our case, both for all electron and BFD pseudopotentials. 
As the degree of exchange mixing is increased, the RS phase becomes more favored. 
For instance, for $\alpha = 60$\% RS has become more stable by $\approx$ 250 (BFD) or 300 (all electron) meV/fu. 
The cross-over occurs around an exchange mixing of $\alpha \approx 10$\%.  
We note the wide variability of relative DFT energy differences predicted for different selections of $\alpha$ in Figure \ref{energyvalpha}. 
For oxides and wide gap semiconductors, although the empirical choice $\alpha = 25$\% in hybrid calculations is motivated from perturbation theory\cite{perturbation} and appears to be quite reasonable in many instances, sometimes tuning of the parameter is required\cite{PhysRevB.76.115109}. 
This sensitivity to simulation parameters renders true quantitative predictions of energy ordering and phase stability challenging within the hybrid DFT framework. 

Regarding the energy differences shown in Fig. \ref{energyvalpha}, our results are consistent with previous results in which several DFT functionals including the Heyd-Scuseria-Ernzerhof screened exchange  hybrid functional (incorrectly) find ZB to be more stable than RS \cite{Schron2010,Peng2013}. 
For instance, using HSE06, the energy difference $E_{ZB}-E_{RS}$ is reported to be -28 meV/fu\cite{Peng2013}.
There are some cautionary notes to be aware of when comparing our results in Fig.~\ref{energyvalpha} to others, however.  
We are using BFD (Hartree-Fock)  rather than DFT pseudopotentials since our primary interest is to generate the best possible QMC description, and not to carry out a DFT study {\it per se}.  
The more localized treatment of the core within Hartree-Fock influences DFT results reported here.  
Also, the lattice constants are fixed to generate the results of Fig. \ref{energyvalpha}, rather than optimized separately for each value of $\alpha$ considered. 
This affects the precise energy differences as well as the ``cross-over" value of $\alpha$. 
Qualitatively, however, Figure \ref{energyvalpha} shows the expected behavior that is consistent with previous results\cite{Peng2013,Schron2010} for this system, and the BFD results appear to be a good starting point for DMC analysis.

\subsection{Density of States}

\begin{figure}
\includegraphics[width=8.5cm]{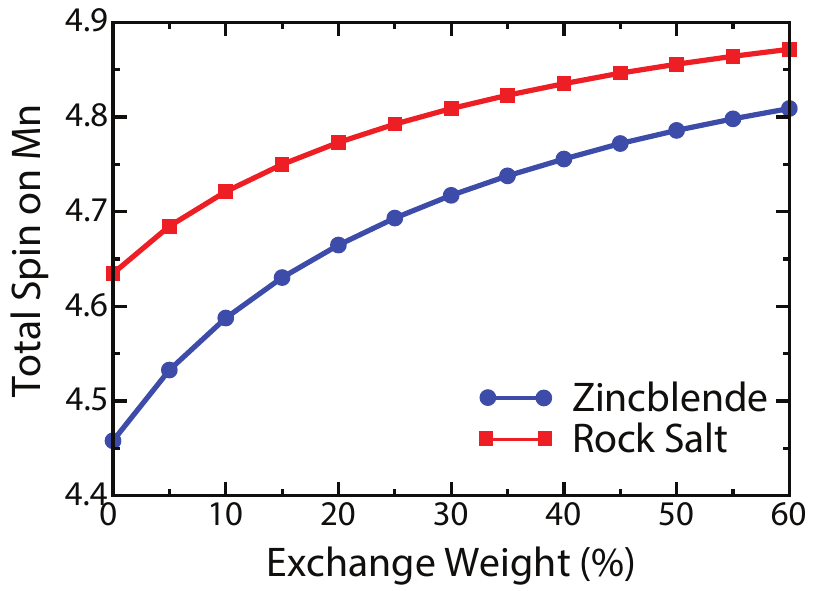}
\caption{\label{spins} (Color online). The total absolute spin on the manganese atoms increases with increasing exchange weight $\alpha$. The spin on the ZB phase is always lower than that of the RS phase.}
\end{figure}

To better understand the DFT trends in Fig. \ref{energyvalpha}, in Fig. \ref{pods} we show the density of states for both phases for different selections of $\alpha$. The black lines are the total DOS, while blue and red respectively indicate states with O $2s$,$2p$ and Mn $3d$ character. The first trend, as expected, is that increasing $\alpha$ widens the band gap in all cases. In addition, $\alpha$ also has an effect on the relative position of the O $2p$ and Mn $3d$ orbitals below the valence band maximum (VBM). For both phases, for $\alpha=0\%$ there are two distinct groups of states: one near the VBM dominated by Mn $3d$ orbitals, and another lower in energy dominated by O $2p$ orbitals. As $\alpha$ increases, the mixing between these sets of orbitals below the VBM increases and the two groups begin to merge; $\alpha$ is essentially a tuning parameter that governs the degree of $p$-$d$ hybridization in the materials. {\it A priori} it is not possible to know which degree of hybridization best captures reality (nor do we expect that sweeping through $\alpha$ will span all possibilities).  
However, to first order increasing $\alpha$ has the effect of canceling the self-interaction error that is present within DFT. Given the trends in Fig. \ref{energyvalpha}, it appears that RS benefits more from this cancellation than ZB ({\it i.e.}, its energy decreases faster as $\alpha$ increases). 

Crystal field theory provides a plausible explanation for why this may be the case. In RS the octahedral field of O surrounding each Mn splits the five degenerate $3d$ orbitals into three lower energy $t_{2g}$ and two higher energy $e_g$ orbitals.  The $t_{2g}$ orbitals are non-bonding, but the $e_g$ orbitals directly overlap and hybridize with the O ligands to form bonding $e_g$ and antibonding $e_g^\ast$ states (see Fig. \ref{pods} for RS for $\alpha = 25$\%). For RS the direct orbital overlap results in a large hybridization and large crystal field splitting between the $t_{2g}$, $e_g$ levels.  By contrast, in the ZB phase the tetrahedral coordination of the Mn $3d$ orbitals results in a splitting of the Mn $d$ orbitals into two lower energy $e$  orbitals and three higher energy $t_2$ orbitals. This time the $e$ orbitals are non-bonding while the $t_2$ orbitals interact with the O ligands (see Fig. \ref{pods} for ZB for $\alpha = 25$\%). The difference is that the $t_2$ orbitals are oriented in between the O orbitals, so the spatial overlap now is less direct. Although the interaction creates bonding and antibonding $t_2$ states, the resulting crystal field splitting $e$, $t_2$ is smaller for ZB.  

We speculate that the direct overlap of orbitals for the RS phase, in contrast to the indirect overlap for ZB, contributes to the difficulty of accurately modeling the RS phase. Greater overlap implies more electrons will occupy the same region in space, which can only be captured by a very good description of electron correlation and exchange. The approximate description of electron correlation in DFT may therefore more adversely affect RS MnO than ZB MnO, causing its energy to be higher than it should be and resulting in the wrong energy ordering.  
 
\subsection{Total Magnetic Moment}

Before proceeding to the DMC results, we also show in Fig. \ref{spins} the total absolute spin on each of the Mn as a function of the exchange weight $\alpha$. For both phases, the total spin increases with increasing $\alpha$. The effect of increasing $\alpha$ diminishes over the domain as the spin approaches 5, the total spin in the ionic limit of the high spin $d^5$ oxide. Furthermore, the total spin on the manganese atoms of ZB is consistently less than those of RS, which is again related to its smaller crystal field splitting. 

\section{Results from DMC Description}

\begin{figure}
\includegraphics[width=8.5cm]{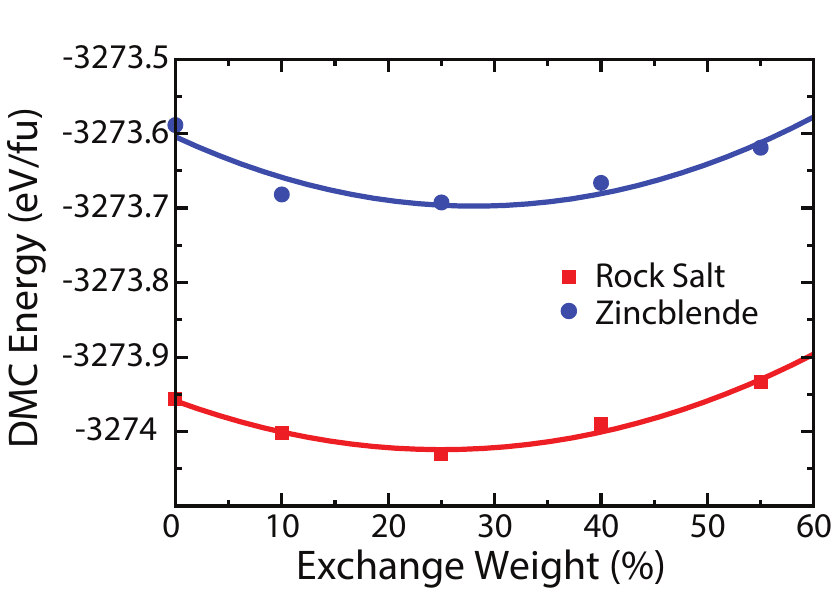} 
\caption{\label{dmcenergyvalpha} (Color online).  
The effect of varying the trial wave function in DMC using different DFT-PBE1$_x$ exchange weights $\alpha$ on the DMC total energies for RS and ZB (error bars are smaller than the marker size). Both phases demonstrate a minimum energy around $\alpha =$ 25\%, and maintain similar relative energies over the domain.}
\end{figure}

\begin{figure}
\includegraphics[width=8.5cm]{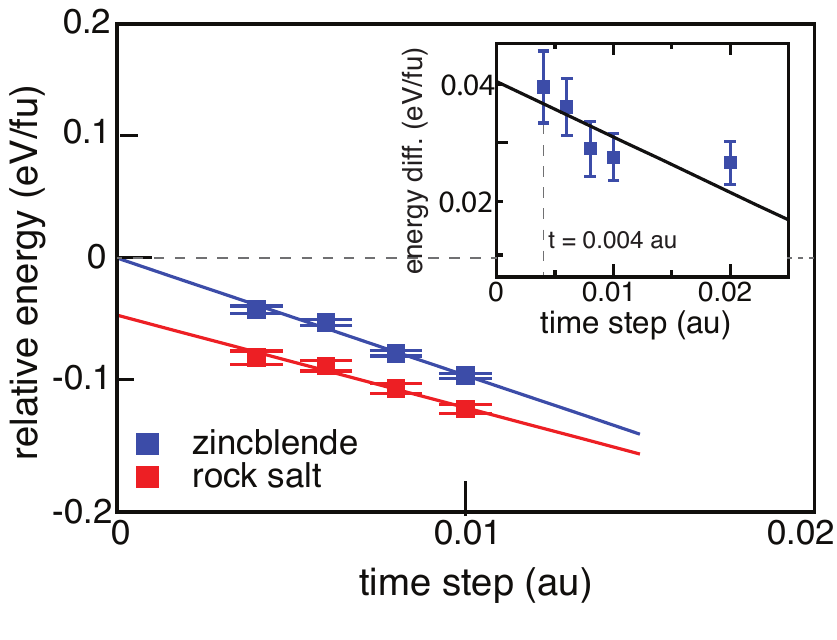} 
\caption{\label{timestep} (Color online).  
The DMC energies of the RS and ZB phase (eV/fu), relative to the extrapolated zero time step DMC energy of the ZB phase, plotted as a function of timestep, for a 4 atom supercell. The inset shows the energy difference $E_{ZB}-E_{RS}$ (eV/fu) vs. time step. } 
\end{figure}

\subsection{Effect of Trial Wave Function}

For the DMC calculations, our first goal is to determine the trial wave function that gives the best description of each phase. Figure \ref{dmcenergyvalpha} shows the total DMC energies for both the RS (red) and ZB (blue) phase as a function of the $\alpha$ used to generate the trial wave function for a 4 atom unit cell.  As DMC is a variational technique, the $\alpha$ that results in the lowest DMC energy gives the best representation of the true nodal surface. Thus, we can use $\alpha$ to vary the nodes of the trial wave function. Although the exact nodal structure is not known, it is expected to sample a wide range since this parameter tunes an important physical quantity: the hybridization between oxygen and manganese.  

For both RS and ZB, a minimum in the DMC energy is observed around $\alpha \approx 25\%$. 
It is interesting to note that this is similar to several other transition metal oxides for which minima in DMC energy tend to occur in a range 15\% $< \alpha <$ 35\% including VO$_2$~\cite{ZhengWagner2015}, FeO\cite{KolorencMitasPRL} , CaCuO$_2$\cite{WagnerAbbamonte}, LaCuO$_4$\cite{WagnerAbbamonte}, and ZnO\cite{YuWagnerErtekinJCP}.
We speculate that $\alpha=25$\% may tend to offer the best description of hybridization between the transition metal $d$ and the O $2p$ orbitals, although we emphasize that this may not always be the case. 
For both phases the overall variation of the total DMC energy is less than 0.15 eV/fu across the full range sampled here, indicating that variations in nodal structure can give rise to total energy differences of roughly this magnitude.  
Nevertheless, the exchange weight that offered the lowest ground state DMC energy within the 4 atom system was used for both polymorphs for all subsequent calculations. This exchange weight was calculated by applying a Bayesian quadratic fit\cite{wagnerJCP2007} to the data of Fig. \ref{dmcenergyvalpha} from which a minimum was determined : $\alpha_{min} \approx 25.0 \pm 0.7\%$ for rock salt, and $\alpha_{min} \approx 28.1 \pm 0.4 \%$ for zincblende.

From the parabolas in Fig. \ref{dmcenergyvalpha}, it appears that when comparing {\it energy differences} between two structures, variations in the nodal structure benefit from a cancellation of errors. 
The DMC energy differences (space between the parabolas for a given $\alpha$) are even less sensitive to changes in the nodal surface that arise from varying $\alpha$ in the trial wave function. The total variation in $(E_{ZB} - E_{RS})$ per fu across the full range of $\alpha$ spanned is now only $0.053 \pm 0.010$ eV (in spite of the 0.3 eV/fu variation exhibited by the DFT starting point calculations). 
We caution that since these DMC results are for 4 atom cells, they suffer from finite size effects and therefore the precise value $(E_{ZB} - E_{RS})$ is not meaningful (we later carry out a full extrapolation of $(E_{ZB} - E_{RS})$ for increasing supercell size).  
Our focus here instead is on the sensitivity of $(E_{ZB} - E_{RS})$ to the trial wave function, which is quite small. It is encouraging that DMC gives consistent results in spite of the large variability of the starting point. 

\subsection{Effect of DMC Timestep}

In diffusion Monte Carlo, a Green's function approach is used to propagate a set of walkers in a 3$N_e$-dimensional space ($N_e$ is the number of electrons),  to statistically sample the many-body wave function. 
The Green's function projector is exact only in the limit of vanishingly small time step, but in practice implementation of DMC requires a finite time step, which introduces an error in the projected energy\cite{anderson75,umrigar93}. 
Therefore, it is important to show that errors in the projected energy due to the finite time step are small, in comparison to the energies of interest. 
In Fig.~\ref{timestep}, we show the DMC energy for RS and ZB (4-atom supercells, twist averaged) as a function of the DMC time step, and the extrapolation of the energy to infinitesimal time step. 
For both phases the dependence of the energy on the time step shows a linear or near-linear dependence, which is expected for sufficiently small time steps. 
For time steps smaller than 0.01 au, for each phase the energy varies within $\approx$ 0.1 eV/fu of the extrapolated value. 

Most importantly, the inset of Fig.~\ref{timestep} shows the energy difference $E_{ZB}-E_{RS}$ (eV/fu) vs. the DMC time step, which is the quantity which we are ultimately interested in resolving. 
This figure shows that {\it energy differences} somewhat benefit from a cancellation of time step errors.
For instance, in the limit of zero time step the energy difference is 0.04(1) eV/fu. 
For a time step of 0.01 au, the computed energy difference instead is around 0.03(1) eV/fu, which indicates a time step error in the energy difference of $\approx$ 0.01(1) eV/fu. 
For a time step of 0.004 au, the computed energy difference is within error bars of the extrapolated energy difference. 
For the remainder of this work, we use a DMC time step of 0.004 a.u.
The uncertainty in $E_{ZB}-E_{RS}$ arising from the time step error here is then less than 0.01 eV/fu, which (as we will demonstrate later) is smaller than the energy difference that we are trying to resolve. 

\subsection{Lattice Constants}

\begin{figure}[h] 
\includegraphics[width=8.5cm]{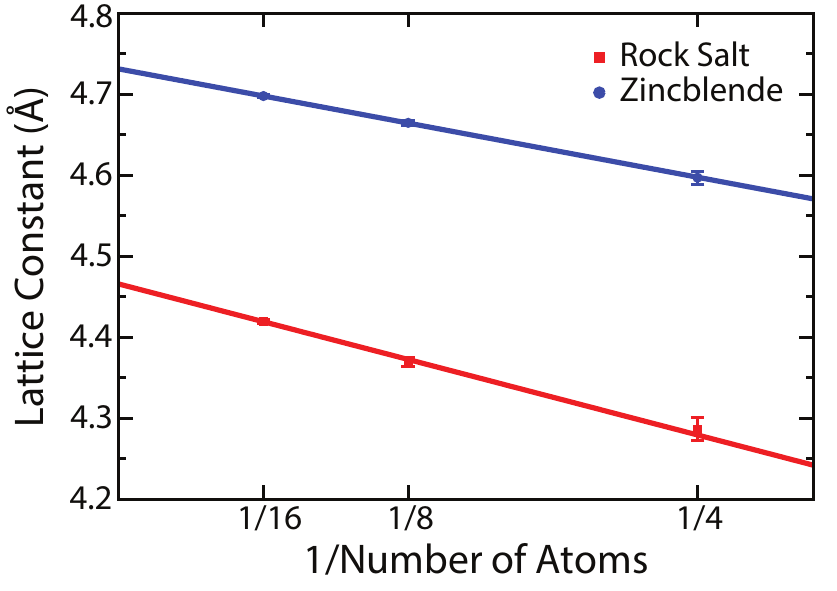} 
\caption{\label{lattconsts} (Color online).  Extrapolated values of lattice constants according to DMC for ZB and RS MnO are $4.73 \pm 0.004$ {\AA} and $4.47 \pm 0.005$ {\AA} respectively. The RS value is within 1\% of experiment 4.43 {\AA}\cite{Roth1958}. The experimental value for ZB is unknown, but the DMC value matches the PBE0 value of $4.73$ {\AA}.  
}
\end{figure}

Since energy differences due to using the wrong lattice constant can be significant when resolving small differences in total energies, it is necessary to  find the optimal lattice constants for both phases within DMC itself. Finding lattice constants in DMC is complicated by the fact that DMC simulations of bulk solids themselves suffer from both one-body and many-body finite size effects. The former are accomodated by twist-averaging, but the latter arise from a spurious correlation between image electrons in the computational domain which typically reduces the total energy\cite{DeLeeuw1980}.  For large enough supercells, the  energy variations scale as $1/V$, where $V$ is the volume of the cell\cite{DeLeeuw1980} (or $1/N$, where $N$ is the number of atoms in the supercell, since $N$ is proportional to $V$). 

Using the optimal $\alpha$ for each phase, we evaluated the total DMC energy of the RS and the ZB phase as a function of the lattice constant $a$, for supercells of size 4, 8, and 16 atoms. Fig. \ref{lattconsts} shows the minimum $a$ obtained for each phase and supercell size. We find that the optimal lattice constant is not the same for different sized supercells but that they increase with increasing supercell size.  To our knowledge, there are no studies of finite size effects on lattice constants within DMC, which would be an interesting avenue for further analysis. From the results in Fig. \ref{lattconsts}, we speculate that many-body finite size effects bias towards smaller lattice constants.  For small supercells, the calculated total energies are artificially low, dominated by the finite size effect. The lattice constant is drawn towards smaller values, which further enhances the stabilizing influence of the spurious image electron correlation.  As the supercell size increases, the finite size effect is reduced and the lattice constants better reflects the true values.

In any case, extrapolating our results to the thermodynamic limit $N \rightarrow \infty$, the lattice constants for the ZB and RS polymorphs of MnO were determined to be $4.730 \pm 0.004$ {\AA} and $4.470 \pm 0.005$ {\AA}. While there is no experimental measurement for the ZB structure, the lattice constant for the RS structure has been previously measured to be 4.43 {\AA}\cite{Roth1958}, demonstrating that in this case our DMC approach can estimate lattice constants to within $\sim$ 1\% of experiment. For the ZB phase the estimated lattice constant matches well the PBE0 value of $4.73$ {\AA} obtained here.  


\subsection{Total Energies and Phase Stability}


\begin{figure}[h] 
\begin{tabular}{l} 
(a) \\
\includegraphics[width=8.5cm]{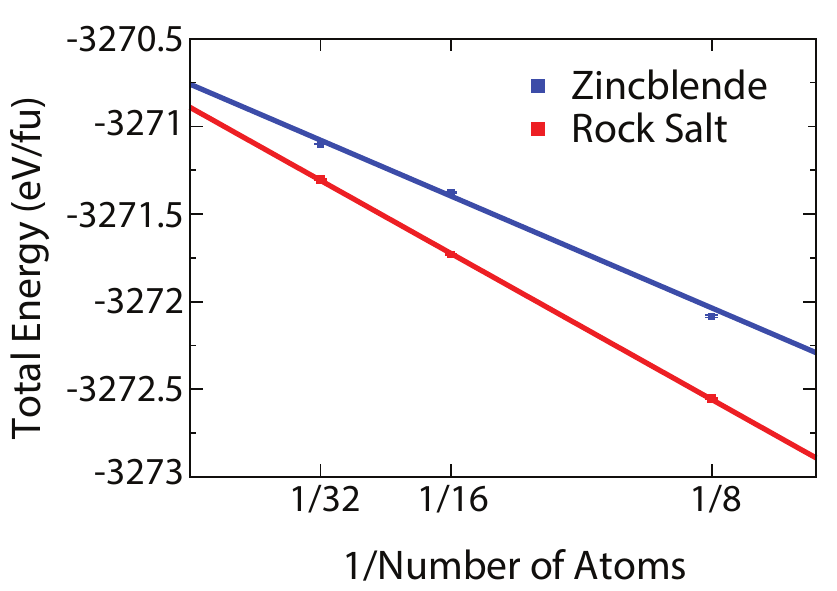} \\
(b) \\
\includegraphics[width=8.5cm]{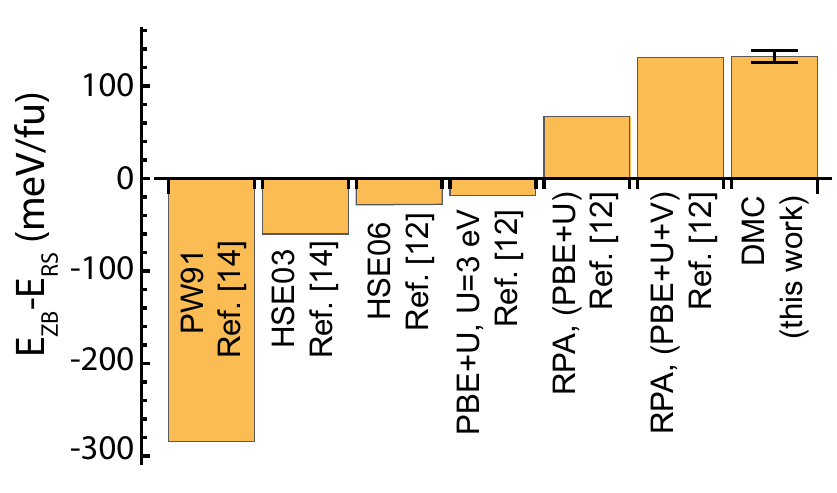} 
\end{tabular} 
\caption{\label{dmcenergies} (Color online).  (a) The extrapolated DMC total energies of the RS and ZB phase; the RS phase is found to be lower in energy by $E_{ZB} - E_{RS} = 132 \pm 6.5$ meV/formula unit. (b) Compiled existing results and current DMC results on the energy ordering of MnO ZB and RS polymorphs.}
\end{figure} 

Finally, using the optimal values of $\alpha$ and the DMC-optimized lattice constants for each phase, we are able to estimate the energy difference for the two phases in the thermodynamic limit. We extrapolated the DMC total energies to the thermodynamic limit using supercells of size $N =$ 8, 16, and 32 atoms.  Our extrapolated DMC results, shown in Fig. \ref{dmcenergies}a, find that the RS phase is more stable than ZB by 132 $\pm$ 6.5 meV/fu. A summary of our DMC results, in comparison to  results obtained using other theories, is presented in Fig. \ref{dmcenergies}b. It is interesting to note the excellent agreement with the estimate pf 131 meV/fu obtained using the random phase approximation to the correlation energy, in conjunction with the application of a Hubbard $U$ as well as a nonlocal external potential $V$ (which are carefully chosen to obtain the correct $p$-$d$ coupling between unoccupied and occupied states, respectively) \cite{Peng2013}. It is encouraging that fixed-node DMC with single determinant wave functions leads to a good description of the basic properties of this highly correlated, antiferromagnetic system. This suggests that this technique can be used for other, similarly complicated materials. 

\subsection{Charge Fluctuations}

\begin{figure}[h] 
\begin{tabular}{l} 
(a) \\
\includegraphics[width=8.5cm]{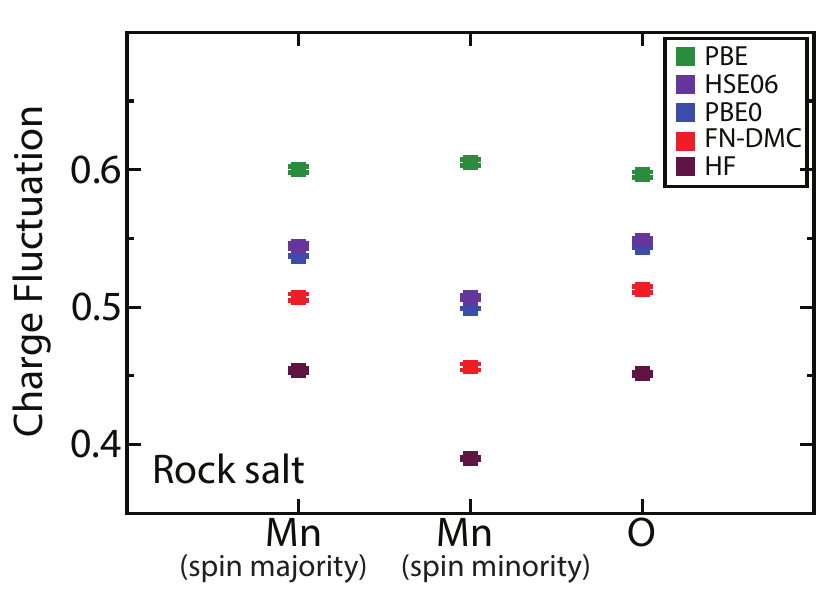} \\
(b) \\
\includegraphics[width=8.5cm]{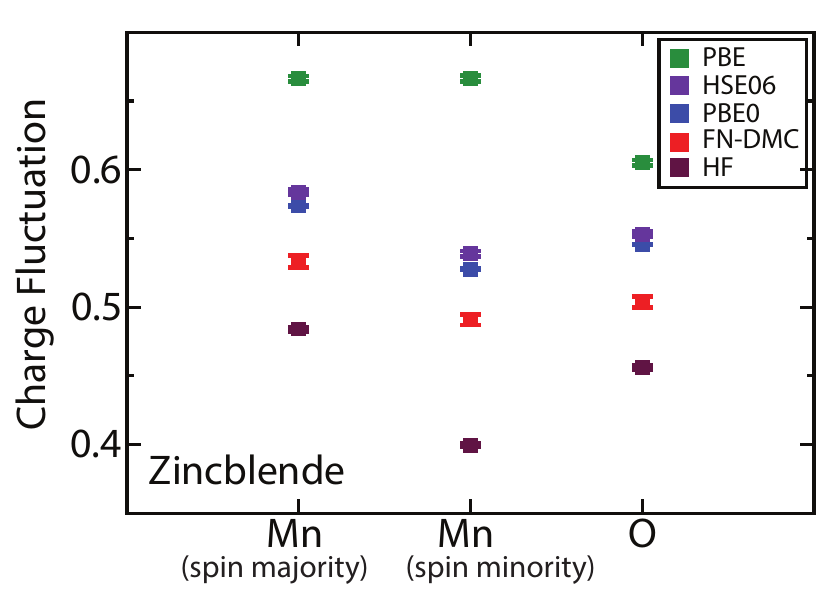} 
\end{tabular} 
\caption{\label{fluctuations} (Color online).  The charge fluctuations $\langle \Psi | (\hat{n}_i - \langle \hat{n}_i \rangle)^2 | \Psi \rangle$ site-resolved onto Mn and O atoms from Slater determinants of PBE, PSE0, HSE06 orbitals, in comparison to DMC and Hartree-Fock for rock salt (top) and zincblende (bottom). The trends demonstrate that both PBE0 and HSE06 improve the description of the materials, bringing the fluctuations closer to that of the DMC values.}
\end{figure}

Since they predict different relative energy ordering, a natural question is ``what changed between the description of the materials in DFT and in DMC?" To provide some insights, in Fig. \ref{fluctuations} we present the site-resolved charge fluctuations, also known as the compressibility, according to both DFT (a Slater determinant composed of the Kohn-Sham orbitals) and DMC. The compressibility is the expectation value $\langle \Psi | ( \hat{n}_i - \left< \hat{n}_i \right>  )^2 | \Psi \rangle$, where $\hat{n}_i$ is the number operator on the Voronoi polyhedron surrounding atomic site $i$. The expectation value is evaluated for a given site by sampling over the DMC configurations of the wave function.  The compressibility  represents the degree to which the number of electrons around a given site fluctuate about the average when the many body wave function is properly sampled. Larger compressibility indicates more fluid charges and delocalized states, while smaller compressibility indicates larger barriers to charge fluctuations and localized states. The charge fluctuations, resolved into majority and minority spins on Mn atoms, are compared for a Slater determinant of DFT (PBE, HSE06, and PBE0) orbitals and our DMC results. We have included the site fluctuations according to Hartree Fock as well for comparison.

According to Figure \ref{fluctuations}, the charge fluctuations vary substantially amongst the different theories. As expected, across the board the fluctuations are largest for PBE (green markers) and smallest for Hartree Fock (brown markers). The DMC results (red markers), presumably the closest to reality, lie somewhere in between. Both PBE0 and HSE06 are observed to improve the description in comparison to PBE, decreasing the charge fluctuations towards the DMC values. It is remarkable that HSE06 and PBE0 both recover the correct qualitative ordering of the fluctuations on the different atomic species.  By contrast, PBE does not get the qualitative ordering correct.  For example for ZB the Mn fluctuations are all larger than the O fluctuations, different from the DMC result. 
Further, Mn in the high spin $d^5$ configuration should have the compressibility of the majority spin higher than that of the minority.  This is properly captured by FN-DMC, HF, and the hybrids.  By contrast, PBE misses this physics entirely both in RS and ZB:  majority and minority spin Mn have similar compressibility. As Fig. \ref{fluctuations} shows, PBE does not describe the localization properly. Ultimately, obtaining the correct energy ordering depends on obtaining a good description of the localization of the states in both phases.  Given the inability to properly describe the localization, PBE cannot be expected to give quantitative information about the relative stability of the materials in question. 


\subsection{Optical Excitation Energies}

Lastly we turn to the DMC calculation of the optical excitation energies for both polymorphs. 
Interest in the polymorphs of MnO has grown recently thanks to computational suggestions that for $d^5$ oxides the zincblende polymorph, although metastable, should have a smaller band gap and a favorable band like hole transport mechanism \cite{Peng2012}. 
Subsequent non-equilibrium growth of Mn$_{1-x}$Zn$_x$O alloys in the wurtzite structure and photo-electrocatalytic device measurements have recently confirmed the predictions \cite{PhysRevX.5.021016}.  
A summary of previously reported band gaps, both from experiment and computation, is given in Table \ref{comparisongaps}.

\begin{table}		
		\begin{tabular}{ l | c }
		Method -- Rock salt & Band Gap (eV) \\ \hline \hline
		 		Conductivity [\onlinecite{FTTa}] & $3.8 - 4.2$ \\ \hline
				Optical Absorption [\onlinecite{FTTb}] & $3.6-3.8$ \\ \hline
				Photoemission spectroscopy [\onlinecite{VanElp1991}] & $3.9 \pm 0.4$  \\ \hline
				Photocurrent measurements [\onlinecite{Usami:1977te}]  & $3.4$ \\ \hline \hline 
		LDA [\onlinecite{Faleev2004}]  & 0.78 \\ \hline
		GGA [\onlinecite{Rodl2009}]  & 0.7 \\ \hline
		HSE03 [\onlinecite{Rodl2009}]  & 2.6 \\ \hline
		LDA + GW [\onlinecite{Faleev2004}]  & 3.5  \\ \hline
		(GGA+U) + GW [\onlinecite{Peng2012}]  & 3.36  \\ \hline
		GGA + GW [\onlinecite{Rodl2009}]  & 1.7 \\ \hline
		HSE03 + GW [\onlinecite{Rodl2009}]  & 3.4 \\ \hline 
		DMC [\onlinecite{Mitas:2010ip}]  & 4.8 $\pm$ 0.2 \\ \hline 
		DMC (QP) (this work) & $4.55 \pm 0.26$ \\  \hline 
		DMC (OG) (this work) & $4.47 \pm 0.16$ \\  
		& \\  
		Method -- Zincblende & Band Gap (eV) \\ \hline \hline
		(GGA+U) + GW (Mn$_{1-x}$Zn$_x$O, $x =0.5 $) [\onlinecite{PhysRevX.5.021016}] &  2.30 \\ \hline \hline
		(GGA+U) + GW [\onlinecite{Peng2012},\onlinecite{PhysRevX.5.021016}] & 2.13, 2.38  \\ \hline
		DMC (QP) (this work) &  $3.55 \pm 0.17$ \\ \hline
		DMC (OP) (this work) & $3.84 \pm 0.14$ \\ 
	\end{tabular}
\caption{\label{comparisongaps} Compiled existing data and current DMC results on the band gaps of MnO ZB and RS polymorphs.}
\end{table}

To these results, we now add the the band gaps of the two phases as obtained from DMC using a procedure that has previously been used successfully \cite{Foulkes:2001td, PhysRevB.57.12140, ErtekinMgO}. 
For both phases, we calculate both the quasiparticle gap (QP) and the optical gap (OG).  
We obtain the quasiparticle gap by calculating the difference between the electron affinity (EA) and the ionization potential (IP): 
\begin{align} 
\label{qp} 
EA &= E(N+1) - E(N) \hspace{2em}, \\
IP &= E(N) - E(N-1) \hspace{2em} \nonumber, \\
QP &= EA - IP  \hspace{2em} \nonumber. 
\end{align} 
Here, $N$ denotes the number of electrons in the neutral solid, $(N+1)$ denotes addition of an electron, and $(N-1)$ denotes removal of an electron. 
The trial wave functions for the DMC calculations to obtain $E(N+1)$, $E(N)$, and $E(N-1)$ in the expressions above are all built from DFT Kohn-Sham orbitals obtained from charge neutral DFT calculations. 
For the case of $E(N+1)$ (or $E(N-1)$), DMC simulations of the charged system is carried out by adding an additional electron to the lowest unfilled orbital (or removing an electron from the highest filled orbital).
Strictly the QP gap should be calculated in the limit $N \rightarrow \infty$; instead we use a 32 atom supercell. 
Both RS and ZB exhibit an indirect gap, but due to zone folding for the 32 atom supercells the gap becomes direct $\Gamma \rightarrow \Gamma$ in both cases. 
Thus we have calculated the QP gap according to Eq. (\ref{qp}) by evaluating the energies at the $\Gamma$ point. 

The optical gap is obtained as 
\begin{equation}
OP = E_{\Gamma \rightarrow \Gamma} - E_o \hspace{2em}. 
\end{equation} 
In this expression $E_o$ denotes the ground state energy and $E_{\Gamma \rightarrow \Gamma} $ denotes the energy of the first optically excited state. 
We estimate the energy difference by evaluating $E_o$ and $E_{\Gamma \rightarrow \Gamma}$ at $\Gamma$, and $E_{\Gamma \rightarrow \Gamma}$ is calculated by promoting an electron from the highest occupied Kohn-Sham orbital at $\Gamma$ to the lowest unoccupied orbital in the construction of the Slater determinant. 
For the OG, once again 32 atom supercells are used. 
For both phases, the OP and the QP obtained in this manner are close, within error bars of each other.  

According to Table \ref{comparisongaps}, compared to previously reported DFT and GW values our DMC results are high (but our gap for the rock salt phase is similar to Kolorenc and Mitas's previous DMC results\cite{Mitas:2010ip}). 
While DMC also predicts that ZB has a lower gap by around 1 eV than RS, the DMC gaps themselves appear to overestimate the experimental values by 0.5 -- 1 eV.  
We suggest several possible reasons for this. 
The first is that finite size effects affect the calculated values. 
A second possibility is that the trial wave functions generated for excited states may not be as good as those generated for the ground state.  
If the nodal structure of the excited state wave function is more complex, then nodal errors will result in an overestimated gap.  

In fact, we note that the case of MnO is particularly challenging for DMC. 
According to the picture from Zaanen, Sawatsky, and Allen \cite{ZaanenSawatskyAllen}, the $3d$ transition metal oxides can be classified as either Mott-Hubbard insulators or charge transfer insulators, based on the degree of $3d$ orbital filling. 
The early $3d$ elements form Mott-Hubbard insulators, for which the gap appears across states of $d$ orbital character (upper and lower Hubbard bands). 
The late $3d$ elements form charge transfer insulators, for which the gap appears across O $2p$ and TM $3d$ states.  
The case of $d^5$ MnO lies just at the transition, suggesting that the VBM has mixed $p-d$ character. 
Since the excited state calculation requires removing an electron from the VBM, the sensitivity to the trial wave function is expected to be particularly strong. 
We suspect that obtaining a better description of the gap depends strongly on generating trial wave functions which more accurately capture the nature of the VBM.  


\section{Conclusions} 
In this work, we have utilized FN-DMC to elucidate the electronic properties and stability of the RS and ZB polymorphs of MnO. 
We find that DMC predicts a ground state energy ordering of these two phases in agreement with experiment without the use of any parameters. 
The energy ordering is insensitive to the choice of the trial wave function, even though different DFT trial wave functions predict very different phase stabilities. 
DMC lattice constants are also in good agreement with experiment. 
Analysis of the site resolved charge fluctuations illustrate some of the primary problems with conventional DFT and show that hybrid functionals make improvements. 
Finally, we find that the DMC calculated band gaps indicate that the tetrahedrally coordinated phase has a lower gap,  but that (within our approach) DMC overestimates the gap according to experiment. 
We attribute this to the challenge of capturing proper description of $p-d$ hybridization in the trial wave functions used in the DMC calculations. 

\acknowledgements
J.A.S. acknowledges the support of a National Science Foundation Graduate Research Fellowship. 
E.E. and L.K.W. acknowledge the support of the National Center for Supercomputing Applications (NCSA) Faculty Fellows program. 
L.K.W was also supported by the U.S. Department of Energy, Office of Science, Office of Advanced Scientific Computing Research, Scientific Discovery through Advanced Computing (SciDAC) program under Award Number FG02-12ER46875.
This research is part of the Blue Waters sustained petascale computing project, which is supported by the National Science 
Foundation (award No. OCI 07-25070 and ACI-1238993) and the state of Illinois. Blue Waters is a joint effort of the University of Illinois at Urbana--Champaign and its National
Center for Supercomputing Applications. 
Also, this research used resources of the Argonne Leadership Computing Facility, which is a DOE Office of Science User Facility supported under Contract DE-AC02-06CH11357.  
Computational resources were also provided by the Illinois Campus Computing cluster. We are very grateful to S. Lany for useful discussions.  

%


\begin{thebibliography}{50}%
\makeatletter
\providecommand \@ifxundefined [1]{%
 \@ifx{#1\undefined}
}%
\providecommand \@ifnum [1]{%
 \ifnum #1\expandafter \@firstoftwo
 \else \expandafter \@secondoftwo
 \fi
}%
\providecommand \@ifx [1]{%
 \ifx #1\expandafter \@firstoftwo
 \else \expandafter \@secondoftwo
 \fi
}%
\providecommand \natexlab [1]{#1}%
\providecommand \enquote  [1]{``#1''}%
\providecommand \bibnamefont  [1]{#1}%
\providecommand \bibfnamefont [1]{#1}%
\providecommand \citenamefont [1]{#1}%
\providecommand \href@noop [0]{\@secondoftwo}%
\providecommand \href [0]{\begingroup \@sanitize@url \@href}%
\providecommand \@href[1]{\@@startlink{#1}\@@href}%
\providecommand \@@href[1]{\endgroup#1\@@endlink}%
\providecommand \@sanitize@url [0]{\catcode `\\12\catcode `\$12\catcode
  `\&12\catcode `\#12\catcode `\^12\catcode `\_12\catcode `\%12\relax}%
\providecommand \@@startlink[1]{}%
\providecommand \@@endlink[0]{}%
\providecommand \url  [0]{\begingroup\@sanitize@url \@url }%
\providecommand \@url [1]{\endgroup\@href {#1}{\urlprefix }}%
\providecommand \urlprefix  [0]{URL }%
\providecommand \Eprint [0]{\href }%
\providecommand \doibase [0]{http://dx.doi.org/}%
\providecommand \selectlanguage [0]{\@gobble}%
\providecommand \bibinfo  [0]{\@secondoftwo}%
\providecommand \bibfield  [0]{\@secondoftwo}%
\providecommand \translation [1]{[#1]}%
\providecommand \BibitemOpen [0]{}%
\providecommand \bibitemStop [0]{}%
\providecommand \bibitemNoStop [0]{.\EOS\space}%
\providecommand \EOS [0]{\spacefactor3000\relax}%
\providecommand \BibitemShut  [1]{\csname bibitem#1\endcsname}%
\let\auto@bib@innerbib\@empty
\bibitem [{\citenamefont {Mott}(1990)}]{Mott1990}%
  \BibitemOpen
  \bibfield  {author} {\bibinfo {author} {\bibfnamefont {N.}~\bibnamefont
  {Mott}},\ }\href {\doibase 10.1016/0022-4596(90)90201-8} {\bibfield
  {journal} {\bibinfo  {journal} {Journal of Solid State Chemistry}\ }\textbf
  {\bibinfo {volume} {88}},\ \bibinfo {pages} {5} (\bibinfo {year}
  {1990})}\BibitemShut {NoStop}%
\bibitem [{\citenamefont {Morin}(1959)}]{Morin1959}%
  \BibitemOpen
  \bibfield  {author} {\bibinfo {author} {\bibfnamefont {F.}~\bibnamefont
  {Morin}},\ }\href {\doibase 10.1103/PhysRevLett.3.34} {\bibfield  {journal}
  {\bibinfo  {journal} {Physical Review Letters}\ }\textbf {\bibinfo {volume}
  {3}},\ \bibinfo {pages} {34} (\bibinfo {year} {1959})}\BibitemShut {NoStop}%
\bibitem [{\citenamefont {Bednorz}\ and\ \citenamefont
  {Mueller}(1986)}]{Bednorz1986}%
  \BibitemOpen
  \bibfield  {author} {\bibinfo {author} {\bibfnamefont {J.~G.}\ \bibnamefont
  {Bednorz}}\ and\ \bibinfo {author} {\bibfnamefont {K.~A.}\ \bibnamefont
  {Mueller}},\ }\href {\doibase 10.1007/BF01303701} {\bibfield  {journal}
  {\bibinfo  {journal} {Zeitschrift fur Physik B Condensed Matter}\ }\textbf
  {\bibinfo {volume} {64}},\ \bibinfo {pages} {189} (\bibinfo {year}
  {1986})}\BibitemShut {NoStop}%
\bibitem [{\citenamefont {Kamihara}\ \emph {et~al.}(2008)\citenamefont
  {Kamihara}, \citenamefont {Watanabe}, \citenamefont {Hirano},\ and\
  \citenamefont {Hosono}}]{Kamihara2008}%
  \BibitemOpen
  \bibfield  {author} {\bibinfo {author} {\bibfnamefont {Y.}~\bibnamefont
  {Kamihara}}, \bibinfo {author} {\bibfnamefont {T.}~\bibnamefont {Watanabe}},
  \bibinfo {author} {\bibfnamefont {M.}~\bibnamefont {Hirano}}, \ and\ \bibinfo
  {author} {\bibfnamefont {H.}~\bibnamefont {Hosono}},\ }\href {\doibase
  10.1021/ja800073m} {\bibfield  {journal} {\bibinfo  {journal} {Journal of the
  American Chemical Society}\ }\textbf {\bibinfo {volume} {130}},\ \bibinfo
  {pages} {3296} (\bibinfo {year} {2008})}\BibitemShut {NoStop}%
\bibitem [{\citenamefont {Urushibara}\ \emph {et~al.}(1995)\citenamefont
  {Urushibara}, \citenamefont {Arima}, \citenamefont {Asamitsu}, \citenamefont
  {Kido},\ and\ \citenamefont {Tokura}}]{Urushibara1995}%
  \BibitemOpen
  \bibfield  {author} {\bibinfo {author} {\bibfnamefont {A.}~\bibnamefont
  {Urushibara}}, \bibinfo {author} {\bibfnamefont {T.}~\bibnamefont {Arima}},
  \bibinfo {author} {\bibfnamefont {A.}~\bibnamefont {Asamitsu}}, \bibinfo
  {author} {\bibfnamefont {G.}~\bibnamefont {Kido}}, \ and\ \bibinfo {author}
  {\bibfnamefont {Y.}~\bibnamefont {Tokura}},\ }\href {\doibase
  10.1103/PhysRevB.51.14103} {\bibfield  {journal} {\bibinfo  {journal}
  {Physical Review B}\ }\textbf {\bibinfo {volume} {51}},\ \bibinfo {pages}
  {14103} (\bibinfo {year} {1995})}\BibitemShut {NoStop}%
\bibitem [{\citenamefont {Lunkenheimer}\ \emph {et~al.}(2010)\citenamefont
  {Lunkenheimer}, \citenamefont {Krohns}, \citenamefont {Riegg}, \citenamefont
  {Ebbinghaus}, \citenamefont {Reller},\ and\ \citenamefont
  {Loidl}}]{Lunkenheimer2010}%
  \BibitemOpen
  \bibfield  {author} {\bibinfo {author} {\bibfnamefont {P.}~\bibnamefont
  {Lunkenheimer}}, \bibinfo {author} {\bibfnamefont {S.}~\bibnamefont
  {Krohns}}, \bibinfo {author} {\bibfnamefont {S.}~\bibnamefont {Riegg}},
  \bibinfo {author} {\bibfnamefont {S.}~\bibnamefont {Ebbinghaus}}, \bibinfo
  {author} {\bibfnamefont {a.}~\bibnamefont {Reller}}, \ and\ \bibinfo {author}
  {\bibfnamefont {a.}~\bibnamefont {Loidl}},\ }\href {\doibase
  10.1140/epjst/e2010-01212-5} {\bibfield  {journal} {\bibinfo  {journal} {The
  European Physical Journal Special Topics}\ }\textbf {\bibinfo {volume}
  {180}},\ \bibinfo {pages} {61} (\bibinfo {year} {2010})}\BibitemShut
  {NoStop}%
\bibitem [{\citenamefont {Peng}\ and\ \citenamefont {Lany}(2012)}]{Peng2012}%
  \BibitemOpen
  \bibfield  {author} {\bibinfo {author} {\bibfnamefont {H.}~\bibnamefont
  {Peng}}\ and\ \bibinfo {author} {\bibfnamefont {S.}~\bibnamefont {Lany}},\
  }\href {\doibase 10.1103/PhysRevB.85.201202} {\bibfield  {journal} {\bibinfo
  {journal} {Physical Review B - Condensed Matter and Materials Physics}\
  }\textbf {\bibinfo {volume} {85}},\ \bibinfo {pages} {201202(R)} (\bibinfo
  {year} {2012})}\BibitemShut {NoStop}%
\bibitem [{\citenamefont {Kanan}\ and\ \citenamefont
  {Carter}(2012)}]{Kanan2012}%
  \BibitemOpen
  \bibfield  {author} {\bibinfo {author} {\bibfnamefont {D.~K.}\ \bibnamefont
  {Kanan}}\ and\ \bibinfo {author} {\bibfnamefont {E.~a.}\ \bibnamefont
  {Carter}},\ }\href {\doibase 10.1021/jp300590d} {\bibfield  {journal}
  {\bibinfo  {journal} {Journal of Physical Chemistry C}\ }\textbf {\bibinfo
  {volume} {116}},\ \bibinfo {pages} {9876} (\bibinfo {year}
  {2012})}\BibitemShut {NoStop}%
\bibitem [{\citenamefont {Toroker}\ and\ \citenamefont
  {Carter}(2013)}]{Toroker2013}%
  \BibitemOpen
  \bibfield  {author} {\bibinfo {author} {\bibfnamefont {M.~C.}\ \bibnamefont
  {Toroker}}\ and\ \bibinfo {author} {\bibfnamefont {E.~a.}\ \bibnamefont
  {Carter}},\ }\href {\doibase 10.1039/c2ta00816e} {\bibfield  {journal}
  {\bibinfo  {journal} {Journal of Materials Chemistry A}\ }\textbf {\bibinfo
  {volume} {1}},\ \bibinfo {pages} {2474} (\bibinfo {year} {2013})}\BibitemShut
  {NoStop}%
\bibitem [{\citenamefont {Gopal}\ \emph {et~al.}(2004)\citenamefont {Gopal},
  \citenamefont {Spaldin},\ and\ \citenamefont {Waghmare}}]{Gopal2004}%
  \BibitemOpen
  \bibfield  {author} {\bibinfo {author} {\bibfnamefont {P.}~\bibnamefont
  {Gopal}}, \bibinfo {author} {\bibfnamefont {N.~a.}\ \bibnamefont {Spaldin}},
  \ and\ \bibinfo {author} {\bibfnamefont {U.~V.}\ \bibnamefont {Waghmare}},\
  }\href {\doibase 10.1103/PhysRevB.70.205104} {\bibfield  {journal} {\bibinfo
  {journal} {Physical Review B - Condensed Matter and Materials Physics}\
  }\textbf {\bibinfo {volume} {70}},\ \bibinfo {pages} {1} (\bibinfo {year}
  {2004})}\BibitemShut {NoStop}%
\bibitem [{\citenamefont {Nam}\ \emph {et~al.}(2012)\citenamefont {Nam},
  \citenamefont {Kim}, \citenamefont {Jo}, \citenamefont {Lee}, \citenamefont
  {Kim}, \citenamefont {Choi}, \citenamefont {Choi}, \citenamefont {Song},\
  and\ \citenamefont {Park}}]{MinNam}%
  \BibitemOpen
  \bibfield  {author} {\bibinfo {author} {\bibfnamefont {K.~M.}\ \bibnamefont
  {Nam}}, \bibinfo {author} {\bibfnamefont {Y.-I.}\ \bibnamefont {Kim}},
  \bibinfo {author} {\bibfnamefont {Y.}~\bibnamefont {Jo}}, \bibinfo {author}
  {\bibfnamefont {S.~M.}\ \bibnamefont {Lee}}, \bibinfo {author} {\bibfnamefont
  {B.~G.}\ \bibnamefont {Kim}}, \bibinfo {author} {\bibfnamefont
  {R.}~\bibnamefont {Choi}}, \bibinfo {author} {\bibfnamefont {S.-I.}\
  \bibnamefont {Choi}}, \bibinfo {author} {\bibfnamefont {H.}~\bibnamefont
  {Song}}, \ and\ \bibinfo {author} {\bibfnamefont {J.~T.}\ \bibnamefont
  {Park}},\ }\href {\doibase 10.1021/ja302440y} {\bibfield  {journal} {\bibinfo
   {journal} {Journal of the American Chemical Society}\ }\textbf {\bibinfo
  {volume} {134}},\ \bibinfo {pages} {8392} (\bibinfo {year} {2012})},\
  \bibinfo {note} {pMID: 22563802},\ \Eprint
  {http://arxiv.org/abs/http://dx.doi.org/10.1021/ja302440y}
  {http://dx.doi.org/10.1021/ja302440y} \BibitemShut {NoStop}%
\bibitem [{\citenamefont {Peng}\ and\ \citenamefont {Lany}(2013)}]{Peng2013}%
  \BibitemOpen
  \bibfield  {author} {\bibinfo {author} {\bibfnamefont {H.}~\bibnamefont
  {Peng}}\ and\ \bibinfo {author} {\bibfnamefont {S.}~\bibnamefont {Lany}},\
  }\href {\doibase 10.1103/PhysRevB.87.174113} {\bibfield  {journal} {\bibinfo
  {journal} {Physical Review B - Condensed Matter and Materials Physics}\
  }\textbf {\bibinfo {volume} {87}},\ \bibinfo {pages} {174113} (\bibinfo
  {year} {2013})}\BibitemShut {NoStop}%
\bibitem [{\citenamefont {Peng}\ \emph {et~al.}(2015)\citenamefont {Peng},
  \citenamefont {Ndione}, \citenamefont {Ginley}, \citenamefont {Zakutayev},\
  and\ \citenamefont {Lany}}]{PhysRevX.5.021016}%
  \BibitemOpen
  \bibfield  {author} {\bibinfo {author} {\bibfnamefont {H.}~\bibnamefont
  {Peng}}, \bibinfo {author} {\bibfnamefont {P.~F.}\ \bibnamefont {Ndione}},
  \bibinfo {author} {\bibfnamefont {D.~S.}\ \bibnamefont {Ginley}}, \bibinfo
  {author} {\bibfnamefont {A.}~\bibnamefont {Zakutayev}}, \ and\ \bibinfo
  {author} {\bibfnamefont {S.}~\bibnamefont {Lany}},\ }\href {\doibase
  10.1103/PhysRevX.5.021016} {\bibfield  {journal} {\bibinfo  {journal} {Phys.
  Rev. X}\ }\textbf {\bibinfo {volume} {5}},\ \bibinfo {pages} {021016}
  (\bibinfo {year} {2015})}\BibitemShut {NoStop}%
\bibitem [{\citenamefont {Schr\"{o}n}\ \emph {et~al.}(2010)\citenamefont
  {Schr\"{o}n}, \citenamefont {R\"{o}dl},\ and\ \citenamefont
  {Bechstedt}}]{Schron2010}%
  \BibitemOpen
  \bibfield  {author} {\bibinfo {author} {\bibfnamefont {A.}~\bibnamefont
  {Schr\"{o}n}}, \bibinfo {author} {\bibfnamefont {C.}~\bibnamefont
  {R\"{o}dl}}, \ and\ \bibinfo {author} {\bibfnamefont {F.}~\bibnamefont
  {Bechstedt}},\ }\href {\doibase 10.1103/PhysRevB.82.165109} {\bibfield
  {journal} {\bibinfo  {journal} {Physical Review B}\ }\textbf {\bibinfo
  {volume} {82}},\ \bibinfo {pages} {165109} (\bibinfo {year}
  {2010})}\BibitemShut {NoStop}%
\bibitem [{\citenamefont {Franchini}\ \emph {et~al.}(2005)\citenamefont
  {Franchini}, \citenamefont {Bayer}, \citenamefont {Podloucky}, \citenamefont
  {Paier},\ and\ \citenamefont {Kresse}}]{Franchini2005}%
  \BibitemOpen
  \bibfield  {author} {\bibinfo {author} {\bibfnamefont {C.}~\bibnamefont
  {Franchini}}, \bibinfo {author} {\bibfnamefont {V.}~\bibnamefont {Bayer}},
  \bibinfo {author} {\bibfnamefont {R.}~\bibnamefont {Podloucky}}, \bibinfo
  {author} {\bibfnamefont {J.}~\bibnamefont {Paier}}, \ and\ \bibinfo {author}
  {\bibfnamefont {G.}~\bibnamefont {Kresse}},\ }\href {\doibase
  10.1103/PhysRevB.72.045132} {\bibfield  {journal} {\bibinfo  {journal} {Phys.
  Rev. B}\ }\textbf {\bibinfo {volume} {72}},\ \bibinfo {pages} {045132}
  (\bibinfo {year} {2005})}\BibitemShut {NoStop}%
\bibitem [{\citenamefont {Towler}\ \emph {et~al.}(1994)\citenamefont {Towler},
  \citenamefont {Allan}, \citenamefont {Harrison}, \citenamefont {Saunders},
  \citenamefont {Mackrodt},\ and\ \citenamefont {Apr\`a}}]{Towler1994}%
  \BibitemOpen
  \bibfield  {author} {\bibinfo {author} {\bibfnamefont {M.~D.}\ \bibnamefont
  {Towler}}, \bibinfo {author} {\bibfnamefont {N.~L.}\ \bibnamefont {Allan}},
  \bibinfo {author} {\bibfnamefont {N.~M.}\ \bibnamefont {Harrison}}, \bibinfo
  {author} {\bibfnamefont {V.~R.}\ \bibnamefont {Saunders}}, \bibinfo {author}
  {\bibfnamefont {W.~C.}\ \bibnamefont {Mackrodt}}, \ and\ \bibinfo {author}
  {\bibfnamefont {E.}~\bibnamefont {Apr\`a}},\ }\href {\doibase
  10.1103/PhysRevB.50.5041} {\bibfield  {journal} {\bibinfo  {journal} {Phys.
  Rev. B}\ }\textbf {\bibinfo {volume} {50}},\ \bibinfo {pages} {5041}
  (\bibinfo {year} {1994})}\BibitemShut {NoStop}%
\bibitem [{\citenamefont {Martin}(2010)}]{MartinText}%
  \BibitemOpen
  \bibfield  {author} {\bibinfo {author} {\bibfnamefont {R.~M.}\ \bibnamefont
  {Martin}},\ }\href@noop {} {\emph {\bibinfo {title} {Electronic structure:
  basic theory and practical methods}}}\ (\bibinfo  {publisher} {Cambridge
  University Press},\ \bibinfo {year} {2010})\BibitemShut {NoStop}%
\bibitem [{\citenamefont {Adamo}\ and\ \citenamefont
  {Barone}(1999)}]{Adamo1999}%
  \BibitemOpen
  \bibfield  {author} {\bibinfo {author} {\bibfnamefont {C.}~\bibnamefont
  {Adamo}}\ and\ \bibinfo {author} {\bibfnamefont {V.}~\bibnamefont {Barone}},\
  }\href {\doibase 10.1063/1.478522} {\bibfield  {journal} {\bibinfo  {journal}
  {The Journal of Chemical Physics}\ }\textbf {\bibinfo {volume} {110}},\
  \bibinfo {pages} {6158} (\bibinfo {year} {1999})}\BibitemShut {NoStop}%
\bibitem [{\citenamefont {Becke}(1993)}]{becke1993new}%
  \BibitemOpen
  \bibfield  {author} {\bibinfo {author} {\bibfnamefont {A.~D.}\ \bibnamefont
  {Becke}},\ }\href@noop {} {\bibfield  {journal} {\bibinfo  {journal} {The
  Journal of Chemical Physics}\ }\textbf {\bibinfo {volume} {98}},\ \bibinfo
  {pages} {1372} (\bibinfo {year} {1993})}\BibitemShut {NoStop}%
\bibitem [{\citenamefont {Heyd}\ \emph {et~al.}(2003)\citenamefont {Heyd},
  \citenamefont {Scuseria},\ and\ \citenamefont {Ernzerhof}}]{heyd2003hybrid}%
  \BibitemOpen
  \bibfield  {author} {\bibinfo {author} {\bibfnamefont {J.}~\bibnamefont
  {Heyd}}, \bibinfo {author} {\bibfnamefont {G.~E.}\ \bibnamefont {Scuseria}},
  \ and\ \bibinfo {author} {\bibfnamefont {M.}~\bibnamefont {Ernzerhof}},\
  }\href@noop {} {\bibfield  {journal} {\bibinfo  {journal} {The Journal of
  Chemical Physics}\ }\textbf {\bibinfo {volume} {118}},\ \bibinfo {pages}
  {8207} (\bibinfo {year} {2003})}\BibitemShut {NoStop}%
\bibitem [{\citenamefont {Petruzielo}\ \emph {et~al.}(2012)\citenamefont
  {Petruzielo}, \citenamefont {Toulouse},\ and\ \citenamefont
  {Umrigar}}]{Petruzielo:2012bl}%
  \BibitemOpen
  \bibfield  {author} {\bibinfo {author} {\bibfnamefont {F.~R.}\ \bibnamefont
  {Petruzielo}}, \bibinfo {author} {\bibfnamefont {J.}~\bibnamefont
  {Toulouse}}, \ and\ \bibinfo {author} {\bibfnamefont {C.~J.}\ \bibnamefont
  {Umrigar}},\ }\href {\doibase 10.1063/1.3697846} {\bibfield  {journal}
  {\bibinfo  {journal} {J. Chem. Phys.}\ }\textbf {\bibinfo {volume} {136}},\
  \bibinfo {pages} {124116} (\bibinfo {year} {2012})}\BibitemShut {NoStop}%
\bibitem [{\citenamefont {Foulkes}\ \emph {et~al.}(2001)\citenamefont
  {Foulkes}, \citenamefont {Mitas}, \citenamefont {Needs},\ and\ \citenamefont
  {Rajagopal}}]{Foulkes:2001td}%
  \BibitemOpen
  \bibfield  {author} {\bibinfo {author} {\bibfnamefont {W.~M.~C.}\
  \bibnamefont {Foulkes}}, \bibinfo {author} {\bibfnamefont {L.}~\bibnamefont
  {Mitas}}, \bibinfo {author} {\bibfnamefont {R.~J.}\ \bibnamefont {Needs}}, \
  and\ \bibinfo {author} {\bibfnamefont {G.}~\bibnamefont {Rajagopal}},\
  }\href@noop {} {\bibfield  {journal} {\bibinfo  {journal} {Rev. Mod. Phys.}\
  }\textbf {\bibinfo {volume} {73}},\ \bibinfo {pages} {33} (\bibinfo {year}
  {2001})}\BibitemShut {NoStop}%
\bibitem [{\citenamefont {Grossman}(2001)}]{Grossman01}%
  \BibitemOpen
  \bibfield  {author} {\bibinfo {author} {\bibfnamefont {J.~C.}\ \bibnamefont
  {Grossman}},\ }\href@noop {} {\bibfield  {journal} {\bibinfo  {journal} {J.
  Chem. Phys.}\ }\textbf {\bibinfo {volume} {117}},\ \bibinfo {pages} {1434}
  (\bibinfo {year} {2001})}\BibitemShut {NoStop}%
\bibitem [{\citenamefont {Wagner}\ \emph {et~al.}(2009)\citenamefont {Wagner},
  \citenamefont {Bajdich},\ and\ \citenamefont {Mitas}}]{Wagner:2009dy}%
  \BibitemOpen
  \bibfield  {author} {\bibinfo {author} {\bibfnamefont {L.~K.}\ \bibnamefont
  {Wagner}}, \bibinfo {author} {\bibfnamefont {M.}~\bibnamefont {Bajdich}}, \
  and\ \bibinfo {author} {\bibfnamefont {L.}~\bibnamefont {Mitas}},\ }\href
  {\doibase 10.1016/j.jcp.2009.01.017} {\bibfield  {journal} {\bibinfo
  {journal} {Journal of Computational Physics}\ }\textbf {\bibinfo {volume}
  {228}},\ \bibinfo {pages} {3390} (\bibinfo {year} {2009})}\BibitemShut
  {NoStop}%
\bibitem [{\citenamefont {Yu}\ \emph {et~al.}()\citenamefont {Yu},
  \citenamefont {Wagner},\ and\ \citenamefont {Ertekin}}]{YuWagnerErtekinJCP}%
  \BibitemOpen
  \bibfield  {author} {\bibinfo {author} {\bibfnamefont {J.}~\bibnamefont
  {Yu}}, \bibinfo {author} {\bibfnamefont {L.~K.}\ \bibnamefont {Wagner}}, \
  and\ \bibinfo {author} {\bibfnamefont {E.}~\bibnamefont {Ertekin}},\ }\href
  {http://arxiv.org/abs/1509.04114} {\bibinfo  {journal} {arXiv:1509.04114v1
  [cond-mat.mtrl-sci]}\ }\BibitemShut {NoStop}%
\bibitem [{\citenamefont {Burkatzki}\ \emph {et~al.}(2007)\citenamefont
  {Burkatzki}, \citenamefont {Filippi},\ and\ \citenamefont
  {Dolg}}]{Burkatzki2007}%
  \BibitemOpen
\bibfield  {journal} {  }\bibfield  {author} {\bibinfo {author} {\bibfnamefont
  {M.}~\bibnamefont {Burkatzki}}, \bibinfo {author} {\bibfnamefont
  {C.}~\bibnamefont {Filippi}}, \ and\ \bibinfo {author} {\bibfnamefont
  {M.}~\bibnamefont {Dolg}},\ }\href {\doibase 10.1063/1.2741534} {\bibfield
  {journal} {\bibinfo  {journal} {Journal of Chemical Physics}\ }\textbf
  {\bibinfo {volume} {126}} (\bibinfo {year} {2007}),\
  10.1063/1.2741534}\BibitemShut {NoStop}%
\bibitem [{\citenamefont {Zheng}\ and\ \citenamefont
  {Wagner}(2015)}]{ZhengWagner2015}%
  \BibitemOpen
  \bibfield  {author} {\bibinfo {author} {\bibfnamefont {H.}~\bibnamefont
  {Zheng}}\ and\ \bibinfo {author} {\bibfnamefont {L.~K.}\ \bibnamefont
  {Wagner}},\ }\href {\doibase 10.1103/PhysRevLett.114.176401} {\bibfield
  {journal} {\bibinfo  {journal} {Phys. Rev. Lett.}\ }\textbf {\bibinfo
  {volume} {114}},\ \bibinfo {pages} {176401} (\bibinfo {year}
  {2015})}\BibitemShut {NoStop}%
\bibitem [{\citenamefont {Devaux}\ \emph {et~al.}(2015)\citenamefont {Devaux},
  \citenamefont {Casula}, \citenamefont {Decremps},\ and\ \citenamefont
  {Sorella}}]{cerium}%
  \BibitemOpen
  \bibfield  {author} {\bibinfo {author} {\bibfnamefont {N.}~\bibnamefont
  {Devaux}}, \bibinfo {author} {\bibfnamefont {M.}~\bibnamefont {Casula}},
  \bibinfo {author} {\bibfnamefont {F.}~\bibnamefont {Decremps}}, \ and\
  \bibinfo {author} {\bibfnamefont {S.}~\bibnamefont {Sorella}},\ }\href
  {\doibase 10.1103/PhysRevB.91.081101} {\bibfield  {journal} {\bibinfo
  {journal} {Phys. Rev. B}\ }\textbf {\bibinfo {volume} {91}},\ \bibinfo
  {pages} {081101} (\bibinfo {year} {2015})}\BibitemShut {NoStop}%
\bibitem [{\citenamefont {Shulenburger}\ \emph {et~al.}(2015)\citenamefont
  {Shulenburger}, \citenamefont {Mattsson},\ and\ \citenamefont
  {Desjarlais}}]{Shulenburger2015}%
  \BibitemOpen
  \bibfield  {author} {\bibinfo {author} {\bibfnamefont {L.}~\bibnamefont
  {Shulenburger}}, \bibinfo {author} {\bibfnamefont {T.~R.}\ \bibnamefont
  {Mattsson}}, \ and\ \bibinfo {author} {\bibfnamefont {M.~P.}\ \bibnamefont
  {Desjarlais}},\ }\href {http://arxiv.org/abs/1501.03850} {\bibfield
  {journal} {\bibinfo  {journal} {arXiv [physics.chem-ph]}\ }\textbf {\bibinfo
  {volume} {1501.03850}} (\bibinfo {year} {2015})}\BibitemShut {NoStop}%
\bibitem [{\citenamefont {Koloren{\v c}}\ and\ \citenamefont
  {Mitas}(2008)}]{KolorencMitasPRL}%
  \BibitemOpen
  \bibfield  {author} {\bibinfo {author} {\bibfnamefont {J.}~\bibnamefont
  {Koloren{\v c}}}\ and\ \bibinfo {author} {\bibfnamefont {L.}~\bibnamefont
  {Mitas}},\ }\href {\doibase 10.1103/PhysRevLett.101.185502} {\bibfield
  {journal} {\bibinfo  {journal} {Phys. Rev. Lett.}\ }\textbf {\bibinfo
  {volume} {101}},\ \bibinfo {pages} {185502} (\bibinfo {year}
  {2008})}\BibitemShut {NoStop}%
\bibitem [{\citenamefont {Mitas}\ and\ \citenamefont
  {Kolorenc}(2010{\natexlab{a}})}]{Mitas2010}%
  \BibitemOpen
  \bibfield  {author} {\bibinfo {author} {\bibfnamefont {L.}~\bibnamefont
  {Mitas}}\ and\ \bibinfo {author} {\bibfnamefont {J.}~\bibnamefont
  {Kolorenc}},\ }\href {\doibase 10.2138/rmg.2010.71.7} {\bibfield  {journal}
  {\bibinfo  {journal} {Reviews in Mineralogy and Geochemistry}\ }\textbf
  {\bibinfo {volume} {71}},\ \bibinfo {pages} {137} (\bibinfo {year}
  {2010}{\natexlab{a}})}\BibitemShut {NoStop}%
\bibitem [{\citenamefont {Dovesi}\ \emph {et~al.}(2005)\citenamefont {Dovesi},
  \citenamefont {Orlando}, \citenamefont {Civalleri}, \citenamefont {Roetti},
  \citenamefont {Saunders},\ and\ \citenamefont
  {Zicovich-Wilson}}]{dovesi2005crystal}%
  \BibitemOpen
  \bibfield  {author} {\bibinfo {author} {\bibfnamefont {R.}~\bibnamefont
  {Dovesi}}, \bibinfo {author} {\bibfnamefont {R.}~\bibnamefont {Orlando}},
  \bibinfo {author} {\bibfnamefont {B.}~\bibnamefont {Civalleri}}, \bibinfo
  {author} {\bibfnamefont {C.}~\bibnamefont {Roetti}}, \bibinfo {author}
  {\bibfnamefont {V.~R.}\ \bibnamefont {Saunders}}, \ and\ \bibinfo {author}
  {\bibfnamefont {C.~M.}\ \bibnamefont {Zicovich-Wilson}},\ }\href@noop {}
  {\bibfield  {journal} {\bibinfo  {journal} {Zeitschrift f{\"u}r
  Kristallographie}\ }\textbf {\bibinfo {volume} {220}},\ \bibinfo {pages}
  {571} (\bibinfo {year} {2005})}\BibitemShut {NoStop}%
\bibitem [{\citenamefont {Roth}(1958)}]{Roth1958}%
  \BibitemOpen
  \bibfield  {author} {\bibinfo {author} {\bibfnamefont {W.~L.}\ \bibnamefont
  {Roth}},\ }\href {\doibase 10.1103/PhysRev.110.1333} {\bibfield  {journal}
  {\bibinfo  {journal} {Physical Review}\ }\textbf {\bibinfo {volume} {110}},\
  \bibinfo {pages} {1333} (\bibinfo {year} {1958})}\BibitemShut {NoStop}%
\bibitem [{\citenamefont {Perdew}\ \emph {et~al.}(1996)\citenamefont {Perdew},
  \citenamefont {Ernzerhof},\ and\ \citenamefont {Burke}}]{perturbation}%
  \BibitemOpen
  \bibfield  {author} {\bibinfo {author} {\bibfnamefont {J.~P.}\ \bibnamefont
  {Perdew}}, \bibinfo {author} {\bibfnamefont {M.}~\bibnamefont {Ernzerhof}}, \
  and\ \bibinfo {author} {\bibfnamefont {K.}~\bibnamefont {Burke}},\ }\href
  {\doibase http://dx.doi.org/10.1063/1.472933} {\bibfield  {journal} {\bibinfo
   {journal} {The Journal of Chemical Physics}\ }\textbf {\bibinfo {volume}
  {105}},\ \bibinfo {pages} {9982} (\bibinfo {year} {1996})}\BibitemShut
  {NoStop}%
\bibitem [{\citenamefont {Fuchs}\ \emph {et~al.}(2007)\citenamefont {Fuchs},
  \citenamefont {Furthm\"uller}, \citenamefont {Bechstedt}, \citenamefont
  {Shishkin},\ and\ \citenamefont {Kresse}}]{PhysRevB.76.115109}%
  \BibitemOpen
  \bibfield  {author} {\bibinfo {author} {\bibfnamefont {F.}~\bibnamefont
  {Fuchs}}, \bibinfo {author} {\bibfnamefont {J.}~\bibnamefont
  {Furthm\"uller}}, \bibinfo {author} {\bibfnamefont {F.}~\bibnamefont
  {Bechstedt}}, \bibinfo {author} {\bibfnamefont {M.}~\bibnamefont {Shishkin}},
  \ and\ \bibinfo {author} {\bibfnamefont {G.}~\bibnamefont {Kresse}},\ }\href
  {\doibase 10.1103/PhysRevB.76.115109} {\bibfield  {journal} {\bibinfo
  {journal} {Phys. Rev. B}\ }\textbf {\bibinfo {volume} {76}},\ \bibinfo
  {pages} {115109} (\bibinfo {year} {2007})}\BibitemShut {NoStop}%
\bibitem [{\citenamefont {Wagner}\ and\ \citenamefont
  {Abbamonte}(2014)}]{WagnerAbbamonte}%
  \BibitemOpen
  \bibfield  {author} {\bibinfo {author} {\bibfnamefont {L.~K.}\ \bibnamefont
  {Wagner}}\ and\ \bibinfo {author} {\bibfnamefont {P.}~\bibnamefont
  {Abbamonte}},\ }\href {\doibase 10.1103/PhysRevB.90.125129} {\bibfield
  {journal} {\bibinfo  {journal} {Phys. Rev. B}\ }\textbf {\bibinfo {volume}
  {90}},\ \bibinfo {pages} {125129} (\bibinfo {year} {2014})}\BibitemShut
  {NoStop}%
\bibitem [{\citenamefont {Wagner}\ and\ \citenamefont
  {Mitas}(2007)}]{wagnerJCP2007}%
  \BibitemOpen
  \bibfield  {author} {\bibinfo {author} {\bibfnamefont {L.~K.}\ \bibnamefont
  {Wagner}}\ and\ \bibinfo {author} {\bibfnamefont {L.}~\bibnamefont {Mitas}},\
  }\href {\doibase http://dx.doi.org/10.1063/1.2428294} {\bibfield  {journal}
  {\bibinfo  {journal} {The Journal of Chemical Physics}\ }\textbf {\bibinfo
  {volume} {126}},\ \bibinfo {eid} {034105} (\bibinfo {year}
  {2007})}\BibitemShut {NoStop}%
\bibitem [{\citenamefont {Anderson}(1975)}]{anderson75}%
  \BibitemOpen
  \bibfield  {author} {\bibinfo {author} {\bibfnamefont {J.~B.}\ \bibnamefont
  {Anderson}},\ }\href {\doibase http://dx.doi.org/10.1063/1.431514} {\bibfield
   {journal} {\bibinfo  {journal} {The Journal of Chemical Physics}\ }\textbf
  {\bibinfo {volume} {63}},\ \bibinfo {pages} {1499} (\bibinfo {year}
  {1975})}\BibitemShut {NoStop}%
\bibitem [{\citenamefont {Umrigar}\ \emph {et~al.}(1993)\citenamefont
  {Umrigar}, \citenamefont {Nightingale},\ and\ \citenamefont
  {Runge}}]{umrigar93}%
  \BibitemOpen
  \bibfield  {author} {\bibinfo {author} {\bibfnamefont {C.~J.}\ \bibnamefont
  {Umrigar}}, \bibinfo {author} {\bibfnamefont {M.~P.}\ \bibnamefont
  {Nightingale}}, \ and\ \bibinfo {author} {\bibfnamefont {K.~J.}\ \bibnamefont
  {Runge}},\ }\href {\doibase http://dx.doi.org/10.1063/1.465195} {\bibfield
  {journal} {\bibinfo  {journal} {The Journal of Chemical Physics}\ }\textbf
  {\bibinfo {volume} {99}},\ \bibinfo {pages} {2865} (\bibinfo {year}
  {1993})}\BibitemShut {NoStop}%
\bibitem [{\citenamefont {de~Leeuw}\ \emph {et~al.}(1980)\citenamefont
  {de~Leeuw}, \citenamefont {Perram},\ and\ \citenamefont
  {Smith}}]{DeLeeuw1980}%
  \BibitemOpen
  \bibfield  {author} {\bibinfo {author} {\bibfnamefont {S.~W.}\ \bibnamefont
  {de~Leeuw}}, \bibinfo {author} {\bibfnamefont {J.~W.}\ \bibnamefont
  {Perram}}, \ and\ \bibinfo {author} {\bibfnamefont {E.~R.}\ \bibnamefont
  {Smith}},\ }\href {\doibase 10.1098/rspa.1980.0135} {\enquote {\bibinfo
  {title} {{Simulation of Electrostatic Systems in Periodic Boundary
  Conditions. I. Lattice Sums and Dielectric Constants}},}\ } (\bibinfo {year}
  {1980})\BibitemShut {NoStop}%
\bibitem [{\citenamefont {Boss}(1968)}]{FTTa}%
  \BibitemOpen
  \bibfield  {author} {\bibinfo {author} {\bibfnamefont {J.}~\bibnamefont
  {Boss}},\ }\href@noop {} {\bibfield  {journal} {\bibinfo  {journal} {Fiz
  Tverd Tela}\ }\textbf {\bibinfo {volume} {10}},\ \bibinfo {pages} {3082}
  (\bibinfo {year} {1968})}\BibitemShut {NoStop}%
\bibitem [{\citenamefont {Iskenderov}\ \emph {et~al.}(1968)\citenamefont
  {Iskenderov}, \citenamefont {Drabkin}, \citenamefont {Emelyanova},\ and\
  \citenamefont {Ksendzov}}]{FTTb}%
  \BibitemOpen
  \bibfield  {author} {\bibinfo {author} {\bibfnamefont {R.~N.}\ \bibnamefont
  {Iskenderov}}, \bibinfo {author} {\bibfnamefont {I.~A.}\ \bibnamefont
  {Drabkin}}, \bibinfo {author} {\bibfnamefont {L.~T.}\ \bibnamefont
  {Emelyanova}}, \ and\ \bibinfo {author} {\bibfnamefont {Y.~M.}\ \bibnamefont
  {Ksendzov}},\ }\href@noop {} {\bibfield  {journal} {\bibinfo  {journal} {Fiz
  Tverd Tela}\ }\textbf {\bibinfo {volume} {10}},\ \bibinfo {pages} {2573}
  (\bibinfo {year} {1968})}\BibitemShut {NoStop}%
\bibitem [{\citenamefont {{Van Elp}}\ \emph {et~al.}(1991)\citenamefont {{Van
  Elp}}, \citenamefont {Potze}, \citenamefont {Eskes}, \citenamefont {Berger},\
  and\ \citenamefont {Sawatzky}}]{VanElp1991}%
  \BibitemOpen
  \bibfield  {author} {\bibinfo {author} {\bibfnamefont {J.}~\bibnamefont {{Van
  Elp}}}, \bibinfo {author} {\bibfnamefont {R.~H.}\ \bibnamefont {Potze}},
  \bibinfo {author} {\bibfnamefont {H.}~\bibnamefont {Eskes}}, \bibinfo
  {author} {\bibfnamefont {R.}~\bibnamefont {Berger}}, \ and\ \bibinfo {author}
  {\bibfnamefont {G.~a.}\ \bibnamefont {Sawatzky}},\ }\href {\doibase
  10.1103/PhysRevB.44.1530} {\bibfield  {journal} {\bibinfo  {journal}
  {Physical Review B}\ }\textbf {\bibinfo {volume} {44}},\ \bibinfo {pages}
  {1530} (\bibinfo {year} {1991})}\BibitemShut {NoStop}%
\bibitem [{\citenamefont {Usami}\ and\ \citenamefont
  {Masumi}(1977)}]{Usami:1977te}%
  \BibitemOpen
  \bibfield  {author} {\bibinfo {author} {\bibfnamefont {T.}~\bibnamefont
  {Usami}}\ and\ \bibinfo {author} {\bibfnamefont {T.}~\bibnamefont {Masumi}},\
  }\href@noop {} {\bibfield  {journal} {\bibinfo  {journal} {Physica BC}\
  }\textbf {\bibinfo {volume} {86-88B}},\ \bibinfo {pages} {985} (\bibinfo
  {year} {1977})}\BibitemShut {NoStop}%
\bibitem [{\citenamefont {Faleev}\ \emph {et~al.}(2004)\citenamefont {Faleev},
  \citenamefont {{Van Schilfgaarde}},\ and\ \citenamefont
  {Kotani}}]{Faleev2004}%
  \BibitemOpen
  \bibfield  {author} {\bibinfo {author} {\bibfnamefont {S.~V.}\ \bibnamefont
  {Faleev}}, \bibinfo {author} {\bibfnamefont {M.}~\bibnamefont {{Van
  Schilfgaarde}}}, \ and\ \bibinfo {author} {\bibfnamefont {T.}~\bibnamefont
  {Kotani}},\ }\href {\doibase 10.1103/PhysRevLett.93.126406} {\bibfield
  {journal} {\bibinfo  {journal} {Physical Review Letters}\ }\textbf {\bibinfo
  {volume} {93}},\ \bibinfo {pages} {12} (\bibinfo {year} {2004})},\ \Eprint
  {http://arxiv.org/abs/0310677} {arXiv:0310677 [cond-mat]} \BibitemShut
  {NoStop}%
\bibitem [{\citenamefont {R\"{o}dl}\ \emph {et~al.}(2009)\citenamefont
  {R\"{o}dl}, \citenamefont {Fuchs}, \citenamefont {Furthm\"{u}ller},\ and\
  \citenamefont {Bechstedt}}]{Rodl2009}%
  \BibitemOpen
  \bibfield  {author} {\bibinfo {author} {\bibfnamefont {C.}~\bibnamefont
  {R\"{o}dl}}, \bibinfo {author} {\bibfnamefont {F.}~\bibnamefont {Fuchs}},
  \bibinfo {author} {\bibfnamefont {J.}~\bibnamefont {Furthm\"{u}ller}}, \ and\
  \bibinfo {author} {\bibfnamefont {F.}~\bibnamefont {Bechstedt}},\ }\href
  {\doibase 10.1103/PhysRevB.79.235114} {\bibfield  {journal} {\bibinfo
  {journal} {Physical Review B - Condensed Matter and Materials Physics}\
  }\textbf {\bibinfo {volume} {79}},\ \bibinfo {pages} {1} (\bibinfo {year}
  {2009})}\BibitemShut {NoStop}%
\bibitem [{\citenamefont {Mitas}\ and\ \citenamefont
  {Kolorenc}(2010{\natexlab{b}})}]{Mitas:2010ip}%
  \BibitemOpen
  \bibfield  {author} {\bibinfo {author} {\bibfnamefont {L.}~\bibnamefont
  {Mitas}}\ and\ \bibinfo {author} {\bibfnamefont {J.}~\bibnamefont
  {Kolorenc}},\ }\href@noop {} {\bibfield  {journal} {\bibinfo  {journal}
  {Reviews in Mineralogy and Geochemistry}\ }\textbf {\bibinfo {volume} {71}},\
  \bibinfo {pages} {137} (\bibinfo {year} {2010}{\natexlab{b}})}\BibitemShut
  {NoStop}%
\bibitem [{\citenamefont {Williamson}\ \emph {et~al.}(1998)\citenamefont
  {Williamson}, \citenamefont {Hood}, \citenamefont {Needs},\ and\
  \citenamefont {Rajagopal}}]{PhysRevB.57.12140}%
  \BibitemOpen
  \bibfield  {author} {\bibinfo {author} {\bibfnamefont {A.~J.}\ \bibnamefont
  {Williamson}}, \bibinfo {author} {\bibfnamefont {R.~Q.}\ \bibnamefont
  {Hood}}, \bibinfo {author} {\bibfnamefont {R.~J.}\ \bibnamefont {Needs}}, \
  and\ \bibinfo {author} {\bibfnamefont {G.}~\bibnamefont {Rajagopal}},\ }\href
  {\doibase 10.1103/PhysRevB.57.12140} {\bibfield  {journal} {\bibinfo
  {journal} {Phys. Rev. B}\ }\textbf {\bibinfo {volume} {57}},\ \bibinfo
  {pages} {12140} (\bibinfo {year} {1998})}\BibitemShut {NoStop}%
\bibitem [{\citenamefont {Ertekin}\ \emph {et~al.}(2013)\citenamefont
  {Ertekin}, \citenamefont {Wagner},\ and\ \citenamefont
  {Grossman}}]{ErtekinMgO}%
  \BibitemOpen
  \bibfield  {author} {\bibinfo {author} {\bibfnamefont {E.}~\bibnamefont
  {Ertekin}}, \bibinfo {author} {\bibfnamefont {L.~K.}\ \bibnamefont {Wagner}},
  \ and\ \bibinfo {author} {\bibfnamefont {J.~C.}\ \bibnamefont {Grossman}},\
  }\href {\doibase 10.1103/PhysRevB.87.155210} {\bibfield  {journal} {\bibinfo
  {journal} {Phys. Rev. B}\ }\textbf {\bibinfo {volume} {87}},\ \bibinfo
  {pages} {155210} (\bibinfo {year} {2013})}\BibitemShut {NoStop}%
\bibitem [{\citenamefont {Zaanen}\ \emph {et~al.}(1985)\citenamefont {Zaanen},
  \citenamefont {Sawatzky},\ and\ \citenamefont {Allen}}]{ZaanenSawatskyAllen}%
  \BibitemOpen
  \bibfield  {author} {\bibinfo {author} {\bibfnamefont {J.}~\bibnamefont
  {Zaanen}}, \bibinfo {author} {\bibfnamefont {G.~A.}\ \bibnamefont
  {Sawatzky}}, \ and\ \bibinfo {author} {\bibfnamefont {J.~W.}\ \bibnamefont
  {Allen}},\ }\href {\doibase 10.1103/PhysRevLett.55.418} {\bibfield  {journal}
  {\bibinfo  {journal} {Phys. Rev. Lett.}\ }\textbf {\bibinfo {volume} {55}},\
  \bibinfo {pages} {418} (\bibinfo {year} {1985})}\BibitemShut {NoStop}%
\end{thebibliography}

\end{document}